# Nearly-incompressible transverse isotropy (NITI) of cornea elasticity: model and experiments with acoustic micro-tapping OCE

**John J. Pitre Jr.[1*], Mitchell A. Kirby[1*], David S. Li[1,2], Tueng T. Shen[3], Ruikang K. Wang[1,3], Matthew O'Donnell[1], and Ivan Pelivanov[1]**

**The cornea provides the largest refractive power for the human visual system. Its stiffness, along with intraocular pressure (IOP), are linked to several pathologies, including keratoconus and glaucoma. Although mechanical tests can quantify corneal elasticity ex vivo, they cannot be used clinically. Optical coherence elastography (OCE), which launches and tracks shear waves to estimate stiffness, provides an attractive non-contact probe of corneal elasticity. To date, however, OCE studies report corneal moduli around tens of kPa, orders-of-magnitude less than those (few MPa) obtained by tensile/inflation testing. This large discrepancy impedes OCE's clinical adoption. Based on corneal microstructure, we introduce and fully characterize a nearly-incompressible transversally isotropic (NITI) model depicting corneal biomechanics. We show that the cornea must be described by two shear moduli, contrary to current single-modulus models, decoupling tensile and shear responses. We measure both as a function of IOP in ex vivo porcine cornea, obtaining values consistent with both tensile and shear tests. At pressures above 30 mmHg, the model begins to fail, consistent with non-linear changes in cornea at high IOP.**

The human cornea provides a unique combination of structure and function to the visual system, serving as a transparent barrier providing two-thirds of the eye's refractive power[1]. It forms a clear refractive lens because of its structure, consisting of collagen fibrils embedded in a hydrated proteoglycan matrix, and is the primary determinant of visual performance[2]. Indeed, its constituents' mechanical properties regulate shape (Fig. 1a), helping focus light onto the retina[3,4].

Because corneal function depends on stiffness, biomechanical metrics can be valuable to both understand and treat corneal disease. For example, keratoconic corneas (Fig. 1b) are measurably less stiff than healthy ones[5]. This observation led to clinical interventions increasing stiffness (e.g. cornea cross-linking)[6]. In procedures such as LASIK and photorefractive keratometry (PRK), an incision releases stromal tension, inducing structural changes that adjust focusing[4]. This close relationship between mechanics and function defines a clinical need for simple and robust measures of corneal stiffness.

In practice, tissue microstructure is too complex to model directly, so assumptions are made to simplify mechanical descriptions of the cornea. The most common model assumes it is a nearly incompressible, isotropic, and linear elastic solid. For this case, a single elastic parameter, the Young's modulus $E$ (or equivalently the shear modulus $\mu = E/3$), defines stiffness. It has been correlated with a number of pathologies and used to design interventions.[7] Unfortunately, measurements require ex vivo tissue samples loaded under tension or inflation. These destructive methods accurately determine corneal $E$, with reported values in the human (in the low-strain region) of 800 kPa to 4.7 MPa for tensile

---

[1.] University of Washington, Department of Bioengineering, Seattle, Washington, United States;
[2.] University of Washington, Department of Chemical Engineering, Seattle, Washington, United States;
[3.] University of Washington, Department of Ophthalmology, Seattle, Washington, United States

loading[8-12], and 100 kPa to 3 MPa for inflation loading[11,13–15]. Although they provide important information on corneal mechanics, their destructive nature precludes clinical translation. Thus, there is clear need for a reliable, non-contact, and non-invasive method to measure corneal biomechanical properties in vivo.

Optical coherence elastography (OCE) is a promising tool addressing this need[16,17]. This method excites elastic shear waves in the cornea (for example, using an air-puff or acoustic micro-tapping, AμT[18]) and tracks them using optical coherence tomography (OCT) (Fig. 1c). Analyzing the shear wave group velocity or dispersion leads to an estimate of tissue shear modulus $\mu$. OCE studies have reported corneal shear moduli in the range of 1.8 – 52.3 kPa[19,20], in close agreement with values obtained from torsional testing of ex vivo cornea (2.5 – 47.3 kPa)[21–23]. However, both shear-based methods produce moduli differing by 1-2 orders of magnitude from those reported by tensile and inflation tests (assuming isotropy, $E = 3\mu$) (Fig. 1d).

Clearly, cornea exhibits markedly different stiffness under shear versus tensile loading, indicating that it cannot be fully described by the single material parameter $E$. Still, the vast majority of corneal mechanics studies over the past 50 years have reported a single Young's modulus, leading to multiple order-of-magnitude discrepancies between results from the two classes of mechanical tests (tensile and shear). Supplementary Note 1 presents a more complete summary of reported Young's and shear moduli, and how they were measured.

We hypothesize that anisotropy is the primary cause of discrepancies between tensile/inflation and torsional/OCE modulus measurements. Corneal microstructure supports this hypothesis. The stroma contain collagen lamellae running in-plane across its width. They make up approximately 90% of tissue thickness and account for the majority of the cornea's mechanical structure. Lamellae are stacked vertically in approximately 200-500 separate planes, with various levels of complexity along depth[24,25], resulting in anisotropic structure and behavior. High-resolution imaging, combined with reported moduli for different loading schemes, strongly suggest that the cornea is anisotropic.

Our goal is to develop an OCE-based technique for clinically translatable measurements consistent with direct mechanical estimates. To obtain reliable, quantitative measurements of corneal moduli, we must address multiple aspects of mechanical wave propagation considering corneal structure. The cornea's finite thickness, bounded by air on one side and liquid on the other, produces complicated guided wave behavior[16]. Adding tissue anisotropy complicates this further, and estimation of moduli is even more difficult. Solutions are not trivial, especially when tissue anisotropy between tensile and shear loading regimes must be considered.

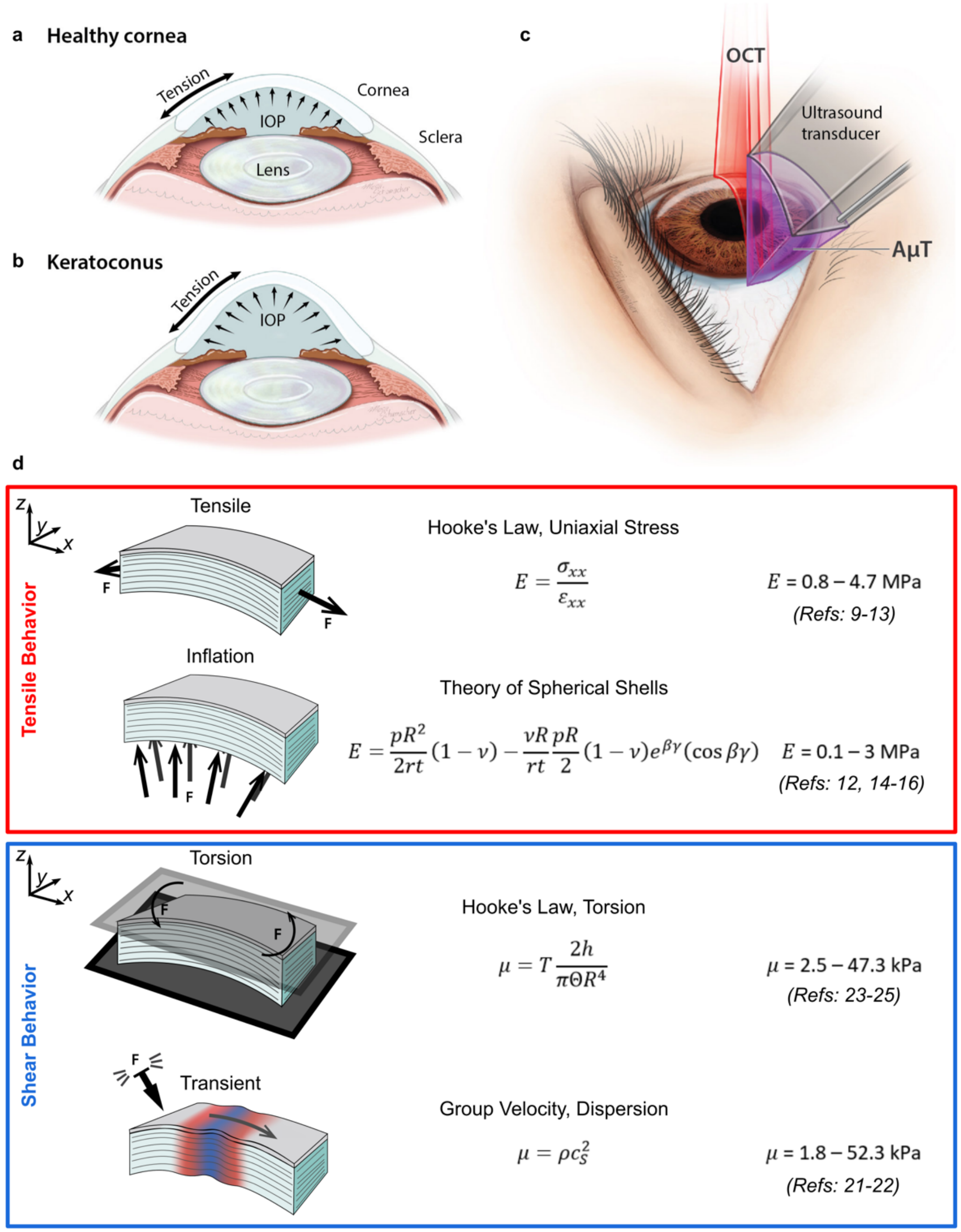

**Fig. 1**. **Relationships between mechanical forces in the cornea and various mechanical testing methods.** Mechanical forces in corneal tissue govern focusing power in (a) healthy and (b) pathological tissue (e.g. keratoconus). (c) Elastic shear waves generated in the cornea with a transient force are tracked using optical coherence tomography to determine corneal mechanical properties without contact. (d) Summary of biomechanical test methods and corneal Young's and shear moduli from the literature. Reported moduli vary by up to four orders of magnitude depending on the loading technique and test configuration, suggesting that isotropic models cannot accurately describe corneal mechanics. In particular, tensile and inflation tests generally agree, as do shear and transient tests. Values listed are for fresh samples (within two weeks), and best effort was taken to report moduli in the low-strain/preload (linear) region.

Here we propose a transversely isotropic (TI) model of the cornea decoupling shear from tensile behavior, thus resolving the apparent paradox in reported biomechanical properties. It intuitively links cornea microstructure to the observed mechanical response. Collagen lamellae contribute to the stiff behavior under tension and inflation (MPa range), while the layered structure allows internal slip producing the softer response of shear and transient tests (kPa range). We compare analytical and simulation results to OCE measurements and demonstrate that the TI model greatly improves quantitative estimates of corneal moduli that agree with ex vivo mechanical tests. These results suggest that clinical stiffness measurements made with OCE may be used to connect the field of non-destructive in vivo testing with the extensive body of ex vivo literature. Bridging this gap is an important step in clinically translating OCE and may help launch large clinical studies in the future.

## Results

**Transversely isotropic model of cornea.** Corneal stroma contain hundreds of vertically stacked collagen lamellae, each 0.2-2.5 μm thick[26] with a preferred collagen orientation. While some in-plane anisotropy has been reported[19,27–30], its magnitude at low intraocular pressure (IOP) suggests that macroscopic behavior can be treated as isotropic in-plane[18,30,31]. Recent second harmonic generation imaging studies show that lamellar orientations are more random than previously suggested[24,32,33], further supporting the assumption of in-plane isotropy. Collagen fiber mechanical properties govern in-plane behavior, while those of the connective tissue matrix govern out-of-plane behavior.

A transversely isotropic (TI) model is the most appropriate given an isotropy plane. It contains five independent elastic constants ($C_{11}$, $C_{12}$, $C_{13}$, $C_{33}$, $C_{44}$) rather than the two (Lamé constants) of isotropic materials. To simplify notation, we adopt the shorthand: $C_{11} = \lambda + 2\mu$, $C_{12} = \lambda$, and $C_{44} = G$. We also assume that the cornea, like most soft tissue, is nearly-incompressible. Specifically, we assume that the medium's internal pressure remains finite as $\lambda \to \infty$ and the dilatation approaches zero[34]. Thus, the stiffness tensor of a nearly-incompressible transversely isotropic (NITI) material can be expressed:

$$C = \begin{bmatrix} \lambda + 2\mu & \lambda & \lambda & & & \\ \lambda & \lambda + 2\mu & \lambda & & & \\ \lambda & \lambda & \lambda + 2\mu & & & \\ & & & G & & \\ & & & & G & \\ & & & & & \mu \end{bmatrix}.$$

The constants λ and μ mimic those in an isotropic solid, with λ enforcing incompressibility and μ defining in-plane shear, tensile, and compressive behavior. Similar to an isotropic material, the Young's modulus is simply related to μ, $E = 3\mu$. An additional shear constant $G$ governs out-of-plane shear and is completely decoupled from $E$.

Uniaxial tensile and inflation tests yield Young's modulus estimates related only to μ. However, shear torsional tests depend only on $G$. In addition, the speed of vertically-polarized bulk shear waves is a function of $G$. This decoupling of normal and shear deformations helps explain the discrepancy between tensile/inflation test modulus estimates (on the order of MPa) and shear/transient estimates (on the order of kPa). Supplementary Note 2 shows how to obtain NITI parameters $\mu$ and $G$ from tensile, inflation, shear, and transient mechanical tests.

**Wave behavior in a bulk NITI medium.** Unlike isotropic materials, which support only two bulk waves, transversely isotropic materials support three – quasi-longitudinal, quasi-shear, and shear. Soft tissues are nearly-incompressible ($\lambda \gg \mu$), implying that the quasi-longitudinal wave speed is nearly constant over all directions. To our knowledge, there are no reports of significant anisotropy for the longitudinal wave speed in cornea. In contrast, quasi-shear and shear wave speeds vary with angle and depend on both $G$ and $\mu$. This has important implications for OCE measurements.

Many dynamic OCE methods track mechanical waves propagating along the air-cornea interface, ignoring liquid loading on the cornea's posterior surface[19,28,35,36]. In other words, the cornea is considered semi-infinite with a simple Rayleigh wave propagating along the surface. The in-plane Rayleigh wave speed can be obtained numerically from the Stroh formalism[37–41]. For materials with $G < \mu$, such as we expect for cornea, the Rayleigh wave speed varies from $c_R = \sqrt{G/\rho}$ in the highly anisotropic limit ($G \ll \mu$) to $c_R = 0.9553\sqrt{G/\rho}$ in the isotropic limit ($G = \mu$) (Supplementary Note 3, Fig. S3.2). Even for varying degrees of anisotropy, Rayleigh wave speed is primarily governed by $G$ and only slightly influenced by $\mu$. This leads to an important conclusion. If the cornea could be considered a semi-infinite TI medium, it would be extremely difficult to determine $\mu$ from OCE measurements. For reference, Supplementary Note 3 derives bulk and Rayleigh wave speeds in a NITI medium.

In reality, the cornea's thickness is typically on the order of the wavelength of propagating mechanical waves considered in dynamic OCE[16,18,42]. As such, corneal thickness and boundary conditions lead to guided waves. As shown below, guided waves provide additional information to help extract both $G$ and $\mu$ from dynamic OCE measurements.

**Guided wave behavior in a bounded NITI layer.** The cornea's bounded geometry produces dispersive guided waves, which must be analyzed in frequency-wavenumber ($\omega$-k) space to quantify elasticity. Partial wave analysis assuming the cornea as a flat isotropic layer bounded by air and water (Fig. 2a) leads to a secular equation describing guided modes[16,20]. Here, we introduce the dispersion relation for a NITI layer bounded by air and water, derived from partial wave solutions to the elastic wave equations satisfying corneal boundary conditions. Supplementary Note 4 provides a full derivation, and functions solving this dispersion relation are provided in Supplementary Software.

While the solution includes an infinite number of quasi-antisymmetric (A) and quasi-symmetric modes (S), only the first two ($A_0$ and $S_0$) are typically captured in OCE. Fig. 2 shows some examples for NITI materials with varying levels of anisotropy ($G$ = 20 kPa, $h$ = 0.55 mm, varying $\mu$). As $\mu$ increases, the $S_0$ mode propagates with higher phase velocity at low frequencies (Fig. 2c). In contrast, the high frequency asymptote of the $A_0$ mode is primarily governed by $G$, with $\mu$ controlling the rate at which the $A_0$ mode approaches its asymptote (Fig. 2d). Thus, the cornea's bounded structure provides a potential avenue to determine both $G$ and $\mu$ using OCE.

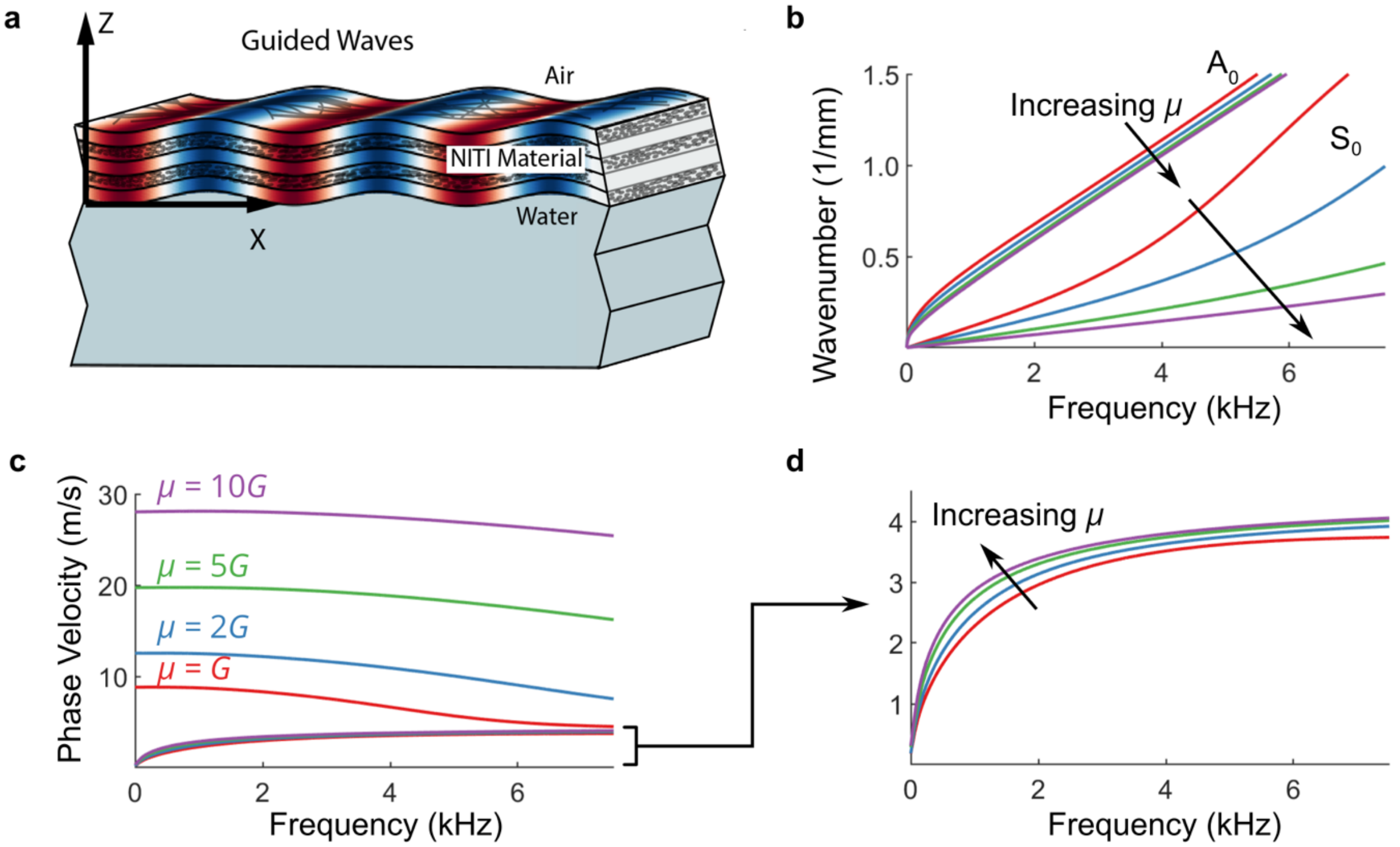

**Fig. 2**. **Analytical solutions for guided wave behavior in a bounded NITI layer.** (a) Propagation of harmonic guided waves in cornea tissue, modeled as a NITI material layer bounded above by air and below by water. Dispersion relations for propagating guided modes are derived from partial wave solutions to the elastic wave equations satisfying corneal boundary conditions in Supplementary Note 4. (b) The effect of increasing anisotropy on fundamental (zero order) $A_0$ and $S_0$ modes ($G$ = 20 kPa, $h$ = 0.55 mm, varying $\mu$) presented in wavenumber-frequency space. The behavior of the $A_0$ mode is primarily governed by $G$, and the $S_0$ mode by $\mu$. (c) The same dispersion curves in phase velocity versus frequency representation. (d) Zoomed view of the $A_0$ mode in phase velocity versus frequency representation - the high-frequency asymptote of the $A_0$ mode depends primarily on $G$, with its low-frequency rate of change governed by $\mu$.

**Finite element model of guided waves in a bounded NITI medium.** Although an analytic solution exists, it is not valid near the mechanical wave excitation source (near field) and does not describe how energy is distributed among guided modes in a NITI layer. To address this, we developed a finite element model using OnScale (OnScale, Redwood City, CA). It contains a NITI layer bounded above by air and below by water with dimensions similar to the cornea. Elastic waves were excited with a spatio-temporally short pressure applied to the top surface, mimicking AμT. Supplementary Note 5 presents a detailed model description, and the OnScale input file is available in Supplementary Software.

We examined temporal changes in the surface velocity field over a range of lateral positions. Unitless surface vibrations illustrate the spatio-temporal behavior of the surface wave field and mimic an OCE measurement (Fig. 3a and 3b). This is referred to as an XT plot. A two-dimensional Fourier transform was then applied to analyze guided wave dispersion. Fig. 3 shows XT plots and corresponding 2D Fourier spectra for increasing $\mu$. In the isotropic limit ($\mu$ = $G$), the XT plot (Fig. 3c) clearly shows multiple interacting guided waves, with $A_0$ and $S_0$ modes both visible in the spectrum (Fig. 3g). As anisotropy increases ($\mu > G$), XT plots (Fig. 3d-3f) change dramatically. Interference between $A_0$ and $S_0$ modes decreases at $\mu = 2G$ and is nearly absent at $\mu = 5G$. Furthermore, as $\mu$ increases, the shape of $A_0$ and $S_0$ modes shift. Energy in the $S_0$ mode also decreases, disappearing almost entirely for $\mu \geq 5G$ (Fig. 3i-3j). These results strongly suggest that if the NITI model is valid, then only one mode, $A_0$, should be visible in cornea. This is in a contrast to an isotropic layer of the same thickness, where multiple high order modes can be observed.

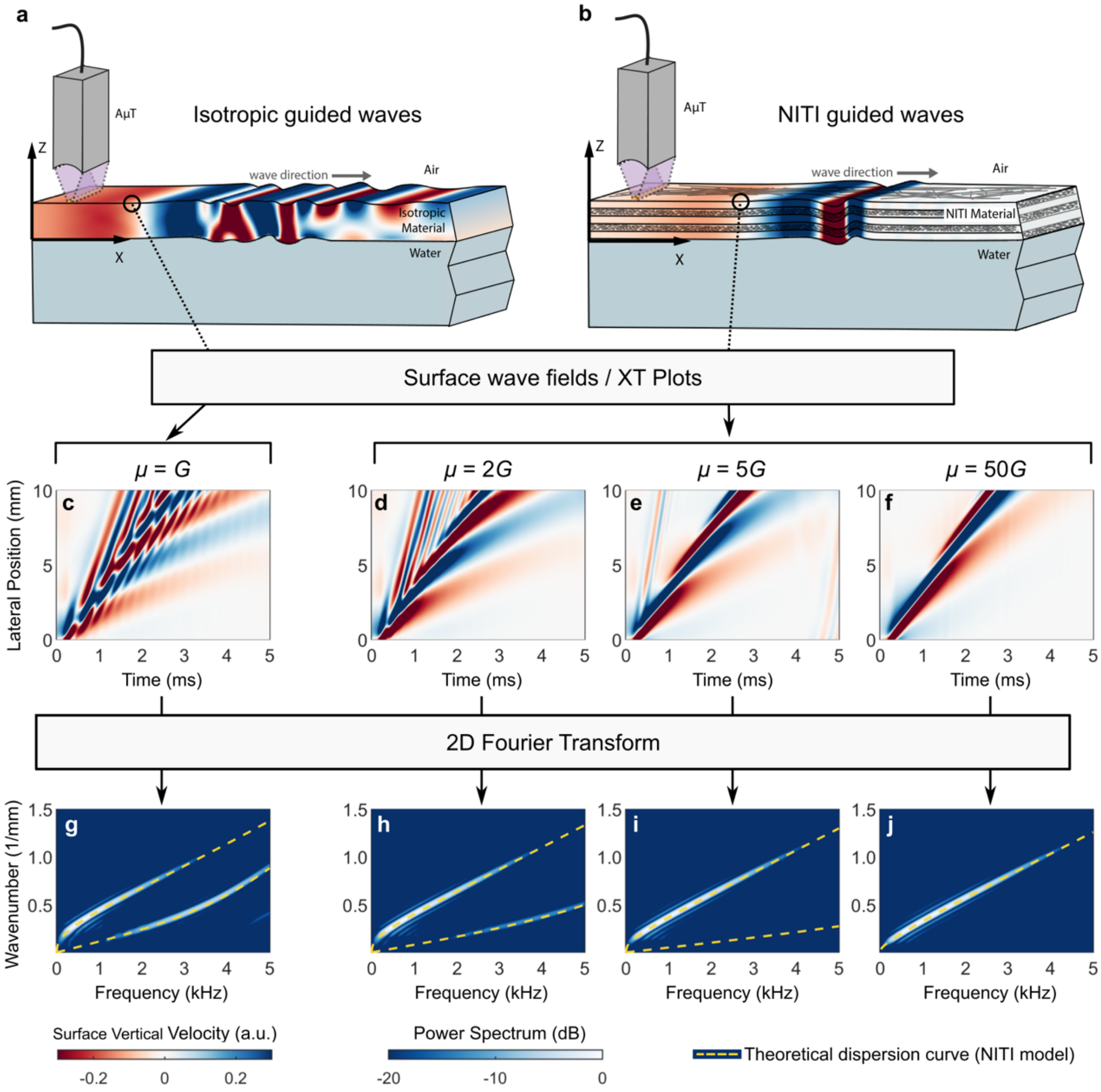

**Fig. 3**. **Numerical solutions for guided wave propagation in a bounded NITI medium.** Guided mode wave fields simulated in OnScale for (a) isotropic and (b) NITI layers. Guided wave excitation was simulated with the AμT line source, closely mimicking experimental conditions (see Supplementary Note 5). Extracting guided mode wave fields from the material surface yields XT plots (c-f) and corresponding 2D Fourier spectra (g-j, presented on a log scale over a 20 dB display dynamic range) for various levels of anisotropy. A 0.55 mm thick NITI layer ($G$ = 20 kPa; $\mu$ = 20 kPa, 40 kPa, 100 kPa, 1 MPa) is considered.

**Elastic modulus estimates with AμT-driven OCE.** A spectral-domain OCT system with a 46.5 kHz effective frame rate, as detailed previously[43], tracked guided waves in isotropic polyvinyl alcohol (PVA) cryogels and porcine cornea, providing experimental measurements to compare with theoretical predictions. A cylindrically-focused 1 MHz air-coupled ultrasound transducer (AμT) provided a spatio-temporally sharp push to the surface of a thin isotropic PVA phantom[43] or freshly excised porcine cornea (n = 6) with IOP incrementally increasing from 5-30 mmHg, generating mechanical waves with bandwidths up to 4 kHz (Methods).

Fig. 4 compares OCE-measured surface velocity fields obtained for an isotropic PVA phantom (Fig. 4a) and porcine cornea at an IOP of 10 mmHg (Fig. 4d). Guided waves are apparent in the isotropic phantom (Fig. 4b). The 2D spectrum of this XT wave field clearly has two guided modes, and fitting an isotropic dispersion relation[16] yields a shear modulus estimate, $\mu_{PVA}$= 14.4 kPa (Fig. 4c, yellow curves).

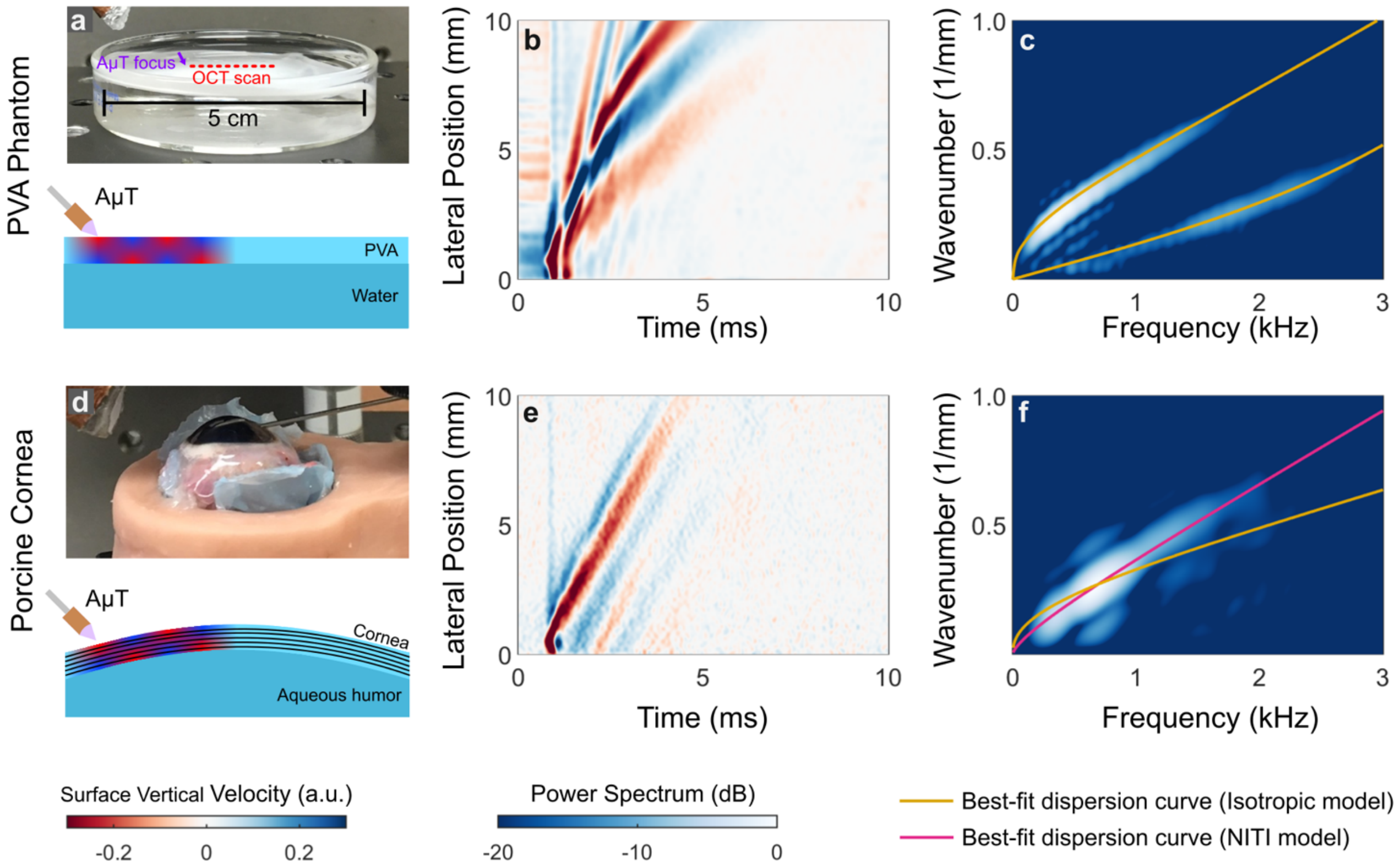

**Fig. 4. AμT-OCE estimation of elastic moduli in an isotropic phantom and ex vivo porcine cornea.** (a) OCE experiments in an isotropic thin PVA phantom bounded above by air and below by water. (b) The guided wave fields excited by AμT are extracted at the phantom surface (XT plot) and a 2D Fourier transform is applied to obtain the frequency-wavenumber spectrum (c, presented on a log scale over a 20 dB display dynamic range). Dispersion curves for an isotropic material are then fit to the spectrum (c, yellow line). The behavior is markedly different from porcine cornea (d) in both the XT plot (e) and wavenumber-frequency spectrum (f). Isotropic and NITI model best fits are shown in (f).

Porcine cornea displays very different behavior, with the wave energy concentrated in a single dispersive mode (Fig. 4e). Two-dimensional spectra highlight differences between PVA and cornea wave fields (Figs. 4c and 4f). Clearly, only the $A_0$ mode is present in porcine cornea (Fig. 4f), as predicted by numerical simulations (Fig. 3).

In porcine cornea, we fit only the $A_0$ mode to both isotropic and NITI dispersion relations. Isotropic fits produced an estimate of the isotropic shear modulus whereas NITI fits produced estimates of both $G$ and $\mu$ ($\mu=E/3$). Using a simplex optimization method (Methods, Supplementary Software), we found dispersion curves that most closely matched the mode structure in 2D spectra. The isotropic model provided a poor fit (Fig. 4f, yellow curve). In contrast, the NITI model closely followed the $A_0$ mode (Fig. 4f, pink curve). This trend was consistent for IOP ranging from 5-20 mmHg (Fig. 5).

Moduli estimates from the NITI model for all corneas and IOPs are summarized in Fig. 6. We observe a multiple order-of-magnitude difference between estimated Young's modulus ($E_{TI} = 3\mu_{TI}$) and the shear modulus $G_{TI}$, consistent with literature values for static tests and OCE measurements. Both moduli increase with IOP over the observed range; however, the order-of-magnitude difference in the moduli remains consistent across all IOP.

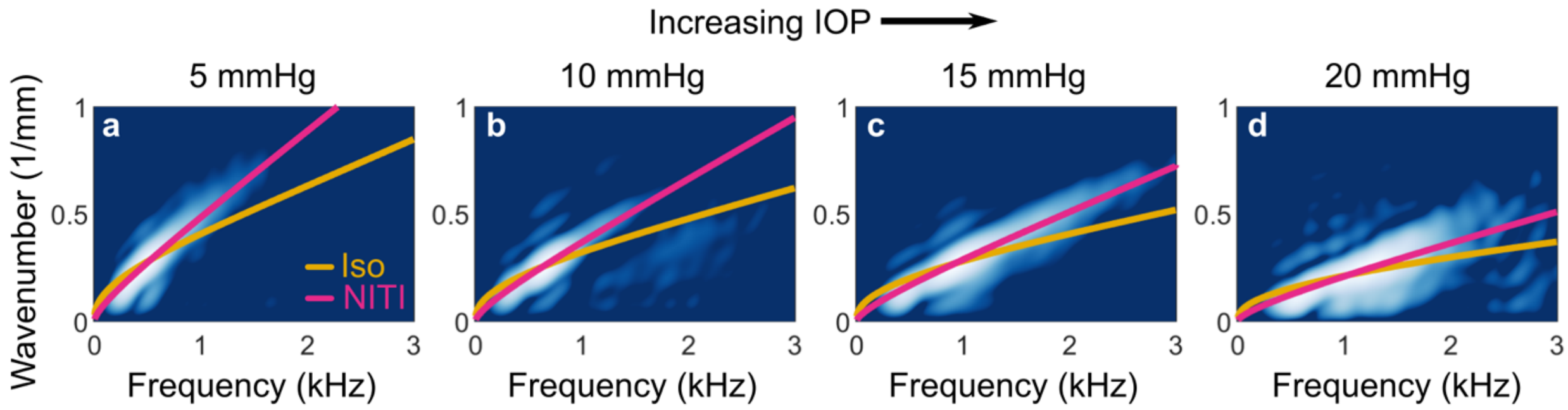

**Fig. 5. 2D Fourier spectra of wave fields generated and tracked with AμT-OCE along the surface of ex vivo porcine cornea at varying intraocular pressure (IOP).** At each IOP, the NITI model (pink line) more closely matches the mode behavior compared to the isotropic model (yellow line). Spectra are shown on a log scale over a 20-dB dynamic range at 5 mmHg IOP (a), 10 mmHg IOP (b), 15 mmHg IOP (c), and 20 mmHg IOP (d).

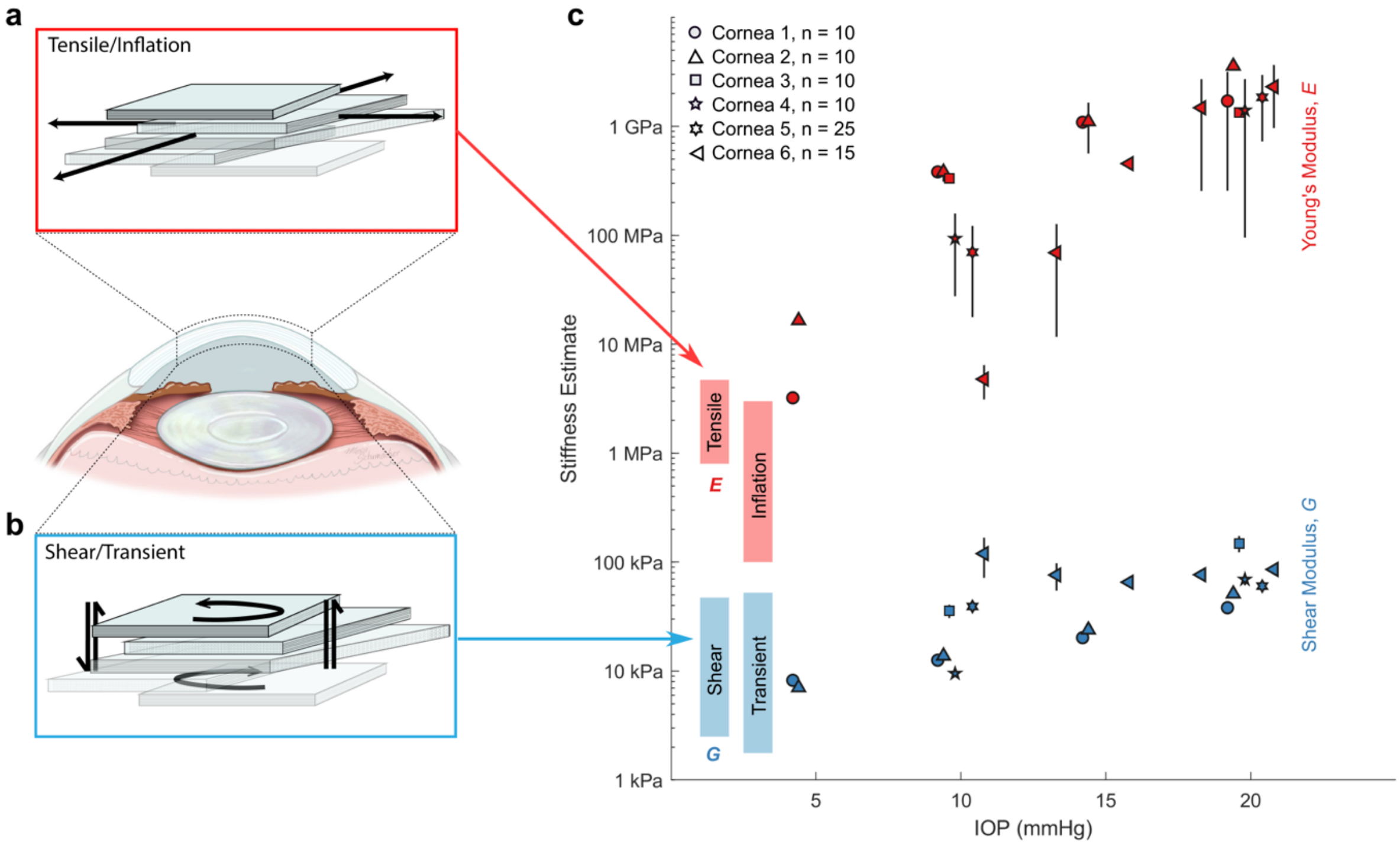

**Fig. 6. NITI elastic moduli estimates (*G* and μ) obtained from AμT-OCE measurements of porcine cornea.** Corneal mechanical response can be divided into (a) tensile/inflation and (b) shear/transient, which can differ by orders of magnitude. (c) Young's modulus (red markers) and shear modulus (blue markers) estimates were obtained from the NITI model for all porcine corneas over a range of IOP. Marker shape corresponds to six individual porcine corneas. For each cornea sample, 10-25 independent scans were taken at each pressure to minimize system variation. Error bars denote ± one standard deviation. Modulus estimates correspond closely to the range of values reported in the literature for static inflation/tensile tests (red bars) and shear/transient measurements (blue bars).

## Discussion

In dynamic OCE, the cornea is typically considered flat, semi-infinite, and isotropic. These approximations may lead to inaccurate interpretations for real corneal geometry and anatomy. For example, the cornea's finite thickness and bounded structure were ignored until recently, when comprehensive simulations and measurements showed that guided modes cannot be ignored.[42]

A flat isotropic layer is also approximate because the cornea is curved. To see the influence on guided mode structure, we performed numerical simulations in flat and spherical bounded layers. The numerical model was similar to the flat layer one (Methods), except the domain was a curved isotropic layer of thickness $h$ = 0.55 mm with outer radius $R_0$ = 6.5 mm. The axisymmetric solution convolved over a line approximates the AμT source. The wave field and Fourier spectrum along the midline of the propagating wave (measured by OCE-AμT) showed little difference between flat and curved models, suggesting that a flat layer can be a reasonable approximation for cornea. This result is unsurprising as curvature adds propagation speed variations of order $h/R_0$ (< 5%)[44]. Considering the dramatic effect of corneal anisotropy, curvature can be ignored. Supplementary Note 6 provides a detailed description of this analysis, as well as the OnScale input file and MATLAB processing functions to reproduce it in Supplementary Software.

Corneal microstructure and an array of biomechanical studies strongly suggest that the cornea is transversely isotropic rather than purely isotropic. Here we have shown that a NITI model more accurately characterizes elastic waves in dynamic OCE studies of the cornea. In particular, elastic waves measured in cornea and isotropic PVA phantoms produce markedly different wave fields and spectra (Fig. 4), demonstrating that an isotropic model is not appropriate for cornea.

The NITI model is defined by two shear moduli ($G$ and $\mu$), decoupling tensile/inflation responses from shear responses commonly monitored in torsional tests and dynamic OCE measurements. Based on existing literature, the Young's modulus for cornea is expected to be on the order of MPa, while the shear modulus is on the order of kPa. This is not physically possible for isotropic materials.

The Rayleigh wave speed in a NITI material is almost entirely defined by the modulus $G$ (see Supplementary Note 3). However, the cornea's finite thickness produces guided waves depending on both $G$ and $\mu$, allowing both parameters to be estimated. Theoretically, the optimal way to determine $\mu$ is from the phase velocity spectrum of the $S_0$ mode, which is largely defined by $\mu$ (Fig. 2c). Unfortunately, numerical simulations and OCE measurements show that excitation at the air/cornea interface transfers little energy to the $S_0$ mode, making it nearly impossible to detect. The absence of an $S_0$ mode is strong evidence of anisotropy, but also means that the $A_0$ mode alone must be used to evaluate both $G$ and $\mu$.

Using the proposed NITI model, we estimated decoupled Young's ($E_{TI} = 3\mu_{TI}$) and shear ($G_{TI}$ ) moduli from measurements of elastic wave propagation in porcine cornea. Results agree well with literature values, with $E_{TI} > 3$ MPa and $G_{TI}$ in the range of 6-100 kPa depending on the IOP (Fig. 6), and generally support the observed order-of-magnitude difference between the two moduli.

We note that $\mu$ shows greater variance relative to $G$. For high anisotropy ($\mu \gg G$), the $A_0$ mode is increasingly insensitive to $\mu$. Measurement noise inside the elastic wave bandwidth has a greater effect on $\mu$ estimates, and uncertainty increases with increasing $\mu$. Thus, while the multiple order-of-magnitude difference between $G$ and $\mu$ is accurate, determining the true value of $\mu$ requires increasingly higher signal-to-noise ratio as anisotropy increases. This produces large confidence intervals for $\mu$ in our measurements.

It is important to note that the NITI model has limitations. As IOP increases, guided modes change noticeably, particularly at very high IOP. Fig. 7 shows two-dimensional spectra and best-fit dispersion curves for a porcine eye measured over a larger IOP range (5, 15, 25, and 35 mmHg). For lower IOP (5 and 15 mmHg), Fourier spectral peaks follow the general $A_0$ mode shape, and the NITI model closely fits the data (Fig. 7a,b). However, at higher IOP (25 and 35 mmHg), mode shape changes dramatically, and the NITI fit no longer describes it well (Fig. 7c,d).

Two factors must be considered in this high IOP regime – nonlinearity and complex anisotropy. The cornea exhibits nonlinear elasticity that changes at approximately 30 mmHg due to a two-stage deformation process[45]. Deformation is governed by a net matrix response at low strain and collagen stiffness at higher strain. Changes in load-bearing characteristics at high IOP may dramatically change wave propagation and/or induce more complex anisotropic behavior.

Stress-strain testing and polarization-sensitive imaging of collagen alignment suggest that the cornea exhibits a relatively symmetric tensile response at low strain[30]. As strain increases, fiber orientation changes. Transient elastography studies have also observed in-plane anisotropy in the cornea starting at 15-20 mmHg and increasing with IOP[18,19,27,28,31]. This suggests that the cornea becomes more nonlinear and anisotropic at high IOP, and the NITI model no longer adequately describes it. Complex anisotropy models, such as orthotropic or fibril-based models, may be required at high IOP.

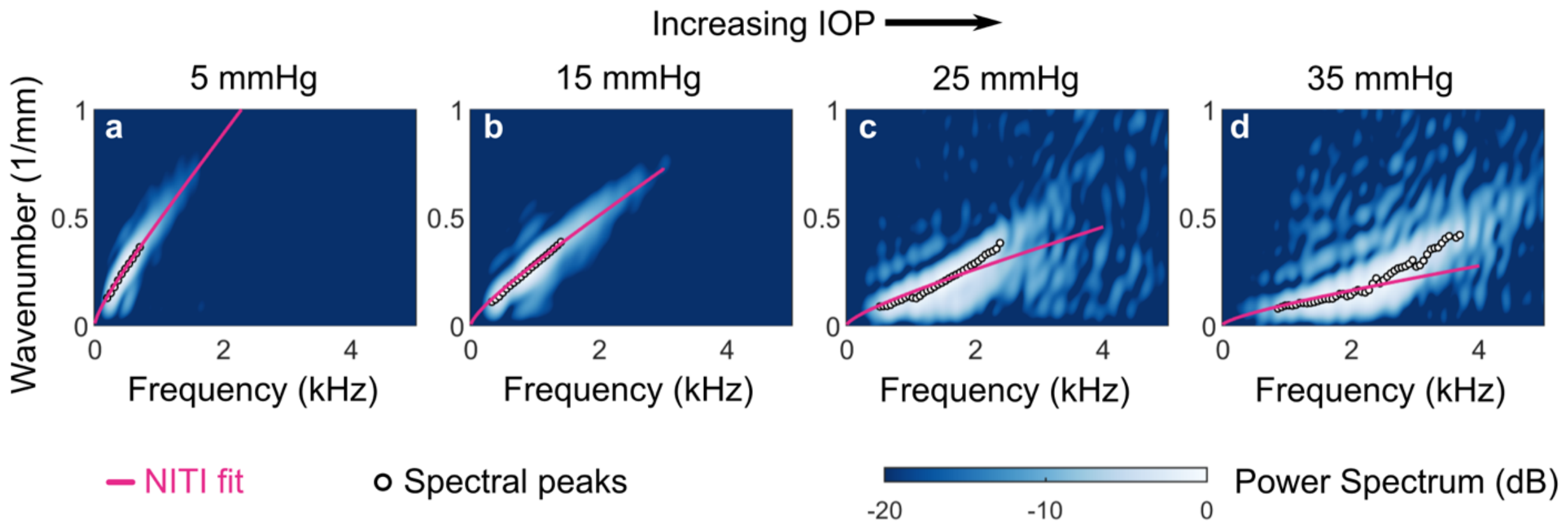

**Fig. 7**. **Nonlinear and anisotropic behavior becomes increasingly complex at high intraocular pressure (IOP).** 2D Fourier spectra of wave fields generated and tracked with AμT-OCE along the surface of porcine cornea at varying intraocular pressure (IOP) show increasingly complex behavior as IOP increases. Spectra are displayed on a log scale over a 20 dB display dynamic range with spectral peaks (circles) and best-fit dispersion curves for NITI model (pink) at 5 mmHg IOP (a), 15 mmHg IOP (b), 25 mmHg IOP (c), and 35 mmHg IOP (d).

Further studies are needed to evaluate the NITI model's clinical utility. As human and porcine cornea differ slightly, it must be tested using OCE measurements on humans. More complex models may also be considered to better estimate corneal stiffness at high IOP. At physiologically relevant IOP, however, we expect the NITI model to perform similarly for human and porcine cornea.

The model's relative simplicity should also facilitate future clinical trials, as it requires a single non-contact measurement, obtained within seconds, to estimate $G$ and $\mu$. While other clinical tools, such as the Ocular Response Analyzer (ORA), have been used clinically to infer both corneal stiffness and IOP[46], sufficient measurement error[47], in addition to patient discomfort[48], hinder widespread use. Further, the relationship between in vivo applanation and corneal Young's Modulus remains unclear[49,50]. The non-contact nature and consistency of AμT-driven OCE strongly support its potential as a practical clinical tool to evaluate corneal elasticity, monitor glaucoma, and study corneal response to ophthalmic interventions.

Non-contact clinical assessment of cornea biomechanics may enable in vivo clinical trials that can provide further insight into the role of biomechanics in cornea function. AμT-OCE can produce accurate maps of mechanical properties, providing reliable, non-contact assessment of corneal biomechanics. Such features make it a potentially valuable tool to evaluate cornea in vivo and to develop future procedures to improve vision by modifying tissue microstructure.

**Methods**

**Analytic solution.** To derive an analytic solution for guided waves in the cornea, we consider an infinite NITI layer of thickness *h* and density ρ bounded above by air and below by water. The stiffness tensor contains material constants λ, μ, and *G*. We assume a plane strain state, consistent with the pseudo-line source generated in our AμT experiments. Here, we briefly overview the derivation for the guided wave solution. Supplementary Note 4 provides a complete derivation. For a NITI material, the elastic wave equations take the following dimensionless form:

$$u_{tt} = \left(\frac{\lambda + 2\mu}{\mu}\right) u_{xx} + \left(\frac{G}{\mu}\right) u_{zz} + \left(\frac{\lambda + G}{\mu}\right) v_{xz},$$

$$v_{tt} = \left(\frac{G}{\mu}\right) v_{xx} + \left(\frac{\lambda + 2\mu}{\mu}\right) v_{zz} + \left(\frac{\lambda + G}{\mu}\right) u_{xz},$$

where *u* and *v* are the *x*- and *z*- components of the displacement, respectively, and subscripts denote partial differentiation. Assuming harmonic plane wave solutions for displacements in the elastic wave equations for the tissue layer and harmonic acoustic wave solutions for the bounding fluid leads to a secular equation that can be solved for the guided wave frequency-wavenumber dispersion relation (Equation S4.16, Supplementary Note 4).

**Numerical simulation.** We developed a finite element numerical model of guided wave propagation in a NITI layer using OnScale (OnScale, Redwood City, CA).[51] Supplementary Note 5 provides a full description of the model. Briefly, we model the cornea as a thin elastic layer of thickness *h* = 0.55 mm and density ρ = 1000 kg/m$^3$, bounded above by air (free surface condition) and below by a layer of water (modeled as an isotropic solid with density 1000 kg/m$^3$, shear wave speed 0 m/s, and a longitudinal wave speed roughly matched to the solid layer). The outer boundaries of the computational domain were set to absorbing conditions. A pressure load was applied to the air-tissue interface with a Gaussian profile in space and super-Gaussian profile in time. The spatial full-width-at-half-max (FWHM) was 600 μm. It was measured with a needle hydrophone (HNC-1000, Onda, Sunnyvale, CA, USA) in air sampling along a 45° line through the AμT transducer focus. The temporal FWHM was 100 μs, also chosen to match AμT experiments.

The computational domain was discretized using linear finite elements on a regular rectangular grid with at least 40 elements per elastic wavelength. Simulations were solved using explicit time stepping, and the vertical velocity component was extracted for analysis, similar to OCE experiments where only this component is available. Velocity data were directional and band-pass filtered using the same processing as OCE experiments to remove reverberations from faster wave components.

**AμT-based OCE to track mechanical waves.** To generate elastic waves, we excited samples with acoustic micro-tapping (AμT), a technique using a cylindrically focused, air-coupled ultrasound transducer to induce a localized radiation force at the sample surface.[18,52] The AμT transducer effectively applied a line load to the surface over a wide region relative to the propagation distance of interest, resulting in approximately planar elastic waves (normal to the OCT image plane). The transducer's full-width-at-half-maximum lateral focus was measured as 420 μm. In the phantom experiment, the spatial push width was approximately 600 μm due to its tilt angle relative to the sample.[18] Due to corneal geometry, the axial focus in the porcine experiment was closer to the theoretically measured axial width of 420μm.[43]

For AμT (as for laser excitation of ultrasound)[53], when an elastic wave is excited by an infinitesimally short (in time) push, the spectral characteristics of the wave are defined by the spatial width of the push. In practice, this pressure confinement can be realized when the push duration is shorter than the shear wave propagation time across the excitation zone. Taking this into account, we utilized a 100 μs pulse duration to generate broadband (up to 4 kHz) mechanical waves and induce tissue displacements on the order of hundreds of nanometers.

The axial particle vibration velocity of propagating mechanical waves was detected using a phase-sensitive frequency-domain OCT (PhS-OCT) system, which has been described in a previous study.[54,55] The sampling rate of the 1024-pixel line-scan InGaAs array was 46.5 kHz, determining the A-line rate of the system (temporal resolution). The optical resolution was approximately 15 μm axially and 24 μm laterally.

To track mechanical wave propagation on the sample surface, an external TTL trigger synchronized the PhS-OCT system with wave excitation for each M-scan. All data were collected in an M-B format in which 512 A-scans are repeated in the same location (M-scan) at 256 different horizontal locations (B-scan) across the imaging plane (dx = 54.7 μm*)*, forming a complete M-B scan (1024 depth × 256 lateral locations × 512 temporal frames) with an effective imaging range of 1.5 mm × 10 mm (axial × lateral). One full M-B scan took 3.66 s.

The resulting three-dimensional dataset was then used to reconstruct the propagating wave based on the OCT-measured local particle vibration velocity. The axial vibration velocity at a given location ($\Delta v_z(x,z,t)$) was obtained from the optical phase difference $\Delta\varphi_{opt}(x,z,t)$ between two consecutive A-line scans at each location using the following equation:[56]

$$\Delta v_z(x,z,t) = \frac{\Delta\varphi_{opt}(x,z,t)\bar{\lambda}}{4\pi\bar{n}f_s^{-1}}$$

where $\bar{\lambda}$ was the center wavelength of the broadband light source, $\bar{n}$ was the refractive index of the medium, and $f_s$ was the sampling frequency. The system was able to reliably detect displacements greater than ≈5 nm.

**Fitting experimental data with the NITI model.** Quantitative moduli estimates in bounded materials requires a method to determine the dispersion relation most closely matching observed guided wave modes. Here, we performed this analysis in the frequency-wavenumber domain using a simplex optimization method (*fminsearch*, MATLAB, MathWorks, Natick, MA). The theoretical solution presented in Supplementary Note 4 acted as the forward model for optimization. A number of physical parameters were considered fixed, including the corneal density (1000 kg/$m^3$), corneal longitudinal wave speed (1540 m/s), and mean corneal thickness (measured from B-mode OCT images). The cornea was bounded from below by water with a density of 1000 kg/$m^3$ and longitudinal wave speed of 1480 m/s. Because we did not observe the $S_0$ mode in corneal measurements, we extracted only the $A_0$ mode from the forward model using a mode-tracing routine (similar to Pavlakovic et. al[57]).

A two-dimensional Fourier transform was applied to OCE-measured surface velocity data to generate a normalized power spectrum. An optimization routine based on the simplex method estimated both shear moduli, $G$ and $\mu$, by fitting the experimentally obtained 2D spectra with the analytic solution (Equation S4.16, Supplementary Note 4). At each iteration, a dispersion relation for the $A_0$ mode was computed for the current iterate ($G_i$, $\mu_i$) based on the forward model. The average power within a small 7-point Gaussian window centered on this dispersion curve was computed, and the algorithm updated iterates of $G$ and $\mu$ to maximize this quantity.

**Isotropic phantom preparation.** A homogenous, isotropic, elastic phantom with controllable mechanical properties was created to experimentally measure wave behavior in a thin plate model. It was fabricated using a similar protocol to that described by Kharine et al.[58] Briefly, polyvinyl alcohol (PVA) (146-186 kDa, >99% hydrolyzed, CAS: 9002-89-5, Sigma-Aldrich Corp., St. Louis, MO, USA) was added to a 4:1 mixture of dimethylsulfoxide (DMSO, CAS: 67-68-5, EMD Millipore Corp) and water at a concentration of 4 wt%. To tune the phantom's optical properties, we added 0.025 wt% titanium dioxide nanoparticles. The solution was covered and stirred at a temperature of 95°C for approximately 1 hour until the PVA was completely dissolved. The solution was degassed in a vacuum chamber to remove any air bubbles before casting in a round mold with a radius of 10 cm. Phantom thickness was controlled by the amount of PVA solution poured into the mold and allowed to settle. The mold was stored at -20°C for at least 12 hours, or until the phantom was completely frozen. The phantom was then thawed at room temperature, completing one freeze-thaw cycle. After casting, phantoms were removed from their molds and placed in a water bath for at least 48 hours to allow the DMSO to diffuse out. Prior to imaging, the PVA phantom was suspended on top of water to force asymmetric boundary conditions similar to those of the cornea.[42]

**Porcine cornea samples.** Porcine eyes were enucleated immediately after death and stored in physiological saline until imaging. All OCE measurements were performed within 1 hour of euthanization. The whole porcine eyeball was placed into a custom-built holder with a hemispherical cup filled with saline-moisturized cotton to provide an in situ environment. The eye globe was oriented cornea side up with the optic axis vertical and aligned with the OCE scanning beam. A 23-gauge needle connected to an infusion reservoir was inserted through the sclera to control intraocular pressure (IOP). The reservoir height was adjusted to maintain IOP between 5 and 40 mmHg.

All studies were carried out in accordance with institutional guidelines and regulations for tissue studies. All experimental protocols followed standard operating procedures established by the University of Washington for the use of animal tissue acquired from an abattoir in research studies.

**Data Availability**

The authors declare that all data from this study are available within the Article and its Supplementary Information. Raw data for the individual measurements are available on reasonable request. In addition, we have included a Supplementary Software Library containing the MATLAB scripts and functions used in this study, as well as the OnScale finite element input files. A detailed description of the functions and scripts is provided in the Supplementary Software Documentation. We also include three example MATLAB data files: (1) Example OCE data from one porcine cornea measurement, (2) example OnScale results for the NITI guided wave model, and (3) example OnScale results for the spherical layer model.

## Acknowledgements

The authors wish to thank Dr. Yak-Nam Wang and the Center for Industrial and Medical Ultrasound at the University of Washington for their assistance in acquiring tissue samples. This work was supported, in part, by NIH grants R01-EY026532, R01-EY024158, R01-EB016034, R01-CA170734, and R01-HL093140, Life Sciences Discovery Fund 3292512, the Coulter Translational Research Partnership Program, an unrestricted grant from the Research to Prevent Blindness, Inc., New York, New York, and the Department of Bioengineering at the University of Washington. M. Kirby was supported by NSF graduate fellowship (No. DGE-1256082). This material was based upon the work supported by the National Science Foundation Graduate Research Fellowship Program under Grant No. DGE-1256082.

## Author Information

**JJP, Jr.** and **MA Kirby** contributed equally to this work.

### Affiliations

[1.] University of Washington, Department of Bioengineering, Seattle, Washington, United States

**John J. Pitre, Jr., Mitchell A. Kirby, David S. Li, Ruikang K. Wang, Matthew O'Donnell, and Ivan Pelivanov**

[2.] University of Washington, Department of Chemical Engineering, Seattle, Washington, United States

**David S. Li**

[3.] University of Washington, Department of Ophthalmology, Seattle, Washington, United States

**Tueng T. Shen, Ruikang K. Wang**

## Contributions

**JJP, Jr.** developed the NITI model to characterize corneal elasticity; performed all analytical studies; designed a finite element model (FEM) to simulate mechanical waves in cornea accounting for its curvature, anisotropy, finite thickness and boundary conditions; ran numerical simulations; developed algorithms to invert wave field data into cornea elastic moduli; processed experimental results; and wrote the paper.

**MAK** conducted OCE experiments, processed experimental data, analyzed the experimental results, and wrote the paper.

**DSL** designed and prepared tissue-mimicking phantoms for OCE experiments.

**TTS** designed the study and wrote the paper.

**RKW** designed the study and wrote the paper.

**MOD** conceived the idea of using a transversely isotropic cornea model with JJP and IP, designed the study, and wrote the paper.

**IP** conceived the idea of using a transversely isotropic cornea model with JJP and MOD, designed the study, and wrote the paper.

## Corresponding Author

Corresponding author is Dr. John J. Pitre, Jr.; e-mail: jpitr@uw.edu

## Ethics Declarations

### Competing Interests

The authors declare they have no competing interests.

# Supplementary Notes

## Nearly-incompressible transverse isotropy (NITI) of cornea elasticity: model and experiments with acoustic micro-tapping OCE

**John J. Pitre Jr.[1*], Mitchell A. Kirby[1*], David S. Li[1,2], Tueng T. Shen[3], Ruikang K. Wang[1,3], Matthew O'Donnell[1], and Ivan Pelivanov[1]**

[1.] University of Washington, Department of Bioengineering, Seattle, Washington, United States
[2.] University of Washington, Department of Chemical Engineering, Seattle, Washington, United States
[3.] University of Washington, Department of Ophthalmology, Seattle, Washington, United States

### Supplementary Note 1. Table of literature-reported mechanical testing results for ex vivo cornea

The following table summarizes the large variability in reported values of elastic moduli for healthy ex vivo cornea in a number of mammalian species. We include moduli values only for the low-strain region, where the corneal stress-strain curve should be nearly linear. Best effort was taken to report values for fresh tissue samples. The reported moduli vary greatly based on the loading method, reconstruction method, and assumed model.

| Tissue Type | Loading Condition | Loading Rate | Reconstruction Method | Assumed Model | Young's Modulus (*E*) | Shear Modulus (*mu*) |
|---|---|---|---|---|---|---|
| Porcine | 15mmHg - 140mmHg Inflation/Displacement | Dynamic | FEM | Hyperelastic nonlinear Ogden material | 300kPa[1] | - |
| Porcine | 15mmHg - 30mmHg Inflation/Air-puff | Dynamic | FEM | Generalized Maxwell viscoelastic | 2.6 MPa[2] | - |
| Porcine | 15mmHg Inflation/Air-puff | Dynamic | FEM | Hyperelastic Mooney Rivlin | .99-1.59 MPa[3] | - |
| Porcine | Tensile | Quasi-static | Stress-strain | - | 1.15-1.93 MPa[3] | - |
| Porcine | 0mmHg - 40mmHg Inflation/Air-puff | Dynamic | Pressure-Deformation | Thin shell | .1-.3 MPa[4] | - |
| Porcine | Tensile | Quasi-static | Stress-strain | - | 3.193±1.589 MPa[4] | - |
| Porcine | .75mmHg- 170mmHg Inflation | Dynamic | Pressure-Deformation | Thin shell | .15-.3 MPa[5] | - |
| Porcine | Tensile | Quasi-static | Stress-strain | - | .3-1.1 MPa[6] | - |
| Porcine | 20mmHg Inflation/Air-Puff OCE | Dynamic | Modified Rayleigh-Lamb Equation | Isotropic Homogenous Viscoelastic | 60 kPa[7] | - |
| Porcine | 15mmHg - 30mmHg Inflation/Air-Puff OCE | Dynamic | Modified Rayleigh-Lamb Equation | Isotropic Homogenous Viscoelastic | 41.8-157 kPa[8] | - |
| Porcine | Compression | Dynamic | Transient compression stress relaxation | Transverse Isotropic Biphasic | 5.61±2.27 kPa (compression)[9] 1.33± .51 MPa (tension)[9] | - |
| Porcine | Compression | Quasi-static | Compressive strain | Transverse Isotropic Biphasic | .65-6.32 MPa[10] | - |
| Porcine | Oscillatory shear | Dynamic (.01-2 Hz) | Shear strain | - | - | 2.5-9 kPa[11] |
| Porcine | 0mmHg - 30mmHg Inflation | Quasi-static | Pressure-Deformation | Linear elastic | 2.29$\pm$1.63MPa[12] | - |
| Porcine | 15mmHg - 35mmHg Inflation/Air-puff | Dynamic | FEM | Linear isotropic viscoelastic | 25.5 MPa (anterior) .85 MPa (posterior)[13] | - |
| Porcine | 15mmHg Inflation/Air-Puff OCE | Dynamic | Group Velocity | Homogenous linear isotropic | 5.9$\pm$.6 kPa[14] | - |
| Porcine | Tensile | Quasi-static | Stress-strain | - | .8-2.2 MPa[15] | - |
| Porcine | Tensile | Quasi-static | Stress-strain | - | 3.70$\pm$ .24 MPa[16] | - |
| Human | Oscillatory shear | Dynamic (.01-2 Hz) | Shear strain | - | - | 3-15 kPa[11] |
| Human | Tensile | Quasi-static | Stress-strain | - | .34-4.1 MPa[17] | - |
| Human | Acoustic Vibration | Dynamic (50-600Hz) | FEM | Linear isotropic viscoelastic | 24.8 kPa (anterior) 19.8 kPa (posterior)[18] | - |
| Human | 15mmHg - 35mmHg Inflation/Air-puff | Dynamic | FEM | Linear isotropic viscoelastic | .71MPa[13] | - |
| Human | Oscillatory shear | Dynamic (.03Hz) | Shear strain | - | - | 38.7$\pm$8.6 kPa[19] |
| Human | 1D Shear | Quasi-static | Shear strain | - | - | 10-90 kPa (Nasal-Temporal) 10-150 kPa (Superior-Inferior)[20] |
| Human | Tensile | Quasi-static | Stress-strain | - | .8-2.6 MPa[15] | - |
| Human | Tensile | Quasi-static | Stress-strain | - | 3.81$\pm$ .40 MPa[16] | - |
| Human | Tensile | Quasi-static | Stress-strain | - | 19.1$\pm$3.5 MPa[21] | - |
| Human | .75mmHg- 160mmHg Inflation | Dynamic | Pressure-Deformation | Thin shell | .25-3 MPa[22] | - |
| Human | .75mmHg- 170mmHg Inflation | Dynamic | Pressure-Deformation | Thin shell | .2-.6 MPa[5] | - |
| Bovine | 1mmHg - 10mmHg Inflation/Piezo-shaker motion MRE | Dynamic (300Hz) | FEM | Linear isotropic | 40-185 kPa[23] | - |
| New Zealand White Rabbit | 15mmHg Inflation/Air-puff | Dynamic | FEM | Nonlinear Hyperelastic Mooney Rivlin plus a Prony-series viscoelastic | 5 MPa[24] | - |
| Rabbit | 15mmHg Inflation/Air-Puff OCE | Dynamic | FEM | Homogenous linear isotropic | 500-800 kPa[25] | - |

**Supplementary Note 2. Behavior of NITI materials under common mechanical tests**

In this study, we show that the nearly incompressible transversely isotropic (NITI) model may explain many of the discrepancies in reported values of corneal Young's modulus. In particular, there is a multiple order-of-magnitude difference between the values reported by tensile/inflation tests and those reported by shear/transient tests. Tensile tests of corneal strips yield values of 800 kPa – 4.7 MPa,[3,4,6,15–17,21] and inflation tests of corneal trephinates yield values of 100 kPa – 3 MPa.[1–5,12,13,22] In contrast, shear (torsional) tests report shear moduli of 2.5 – 47.3 kPa.[11,20,26] This corresponds to a Young's modulus of 7.5 – 142 kPa if one assumes an isotropic model. Likewise, transient methods such as optical coherence elastography (OCE) report Young's moduli in the range 5.3 – 157 kPa,[8,27] again relying on an isotropic model to convert shear modulus to Young's modulus.

The NITI model has a key property that can explain both regimes of reported moduli – namely, its shear and tensile behavior are decoupled. Mechanical tests that probe these behaviors independently will report greatly different Young's moduli. In fact, shear tests will report an independent parameter unrelated to Young's modulus. The shear behavior is governed by an independent modulus, which we refer to as $G$. The tensile behavior is governed by the modulus $\mu$, which can be related to the Young's modulus $E_{TI}$ = 3$\mu$. Here, we demonstrate that the relevant stress-strain relations involved in each test type will depend primarily on one elastic parameter while remaining agnostic to the other. Furthermore, tensile and inflation measurements should yield similar values, as should shear and transient ones.

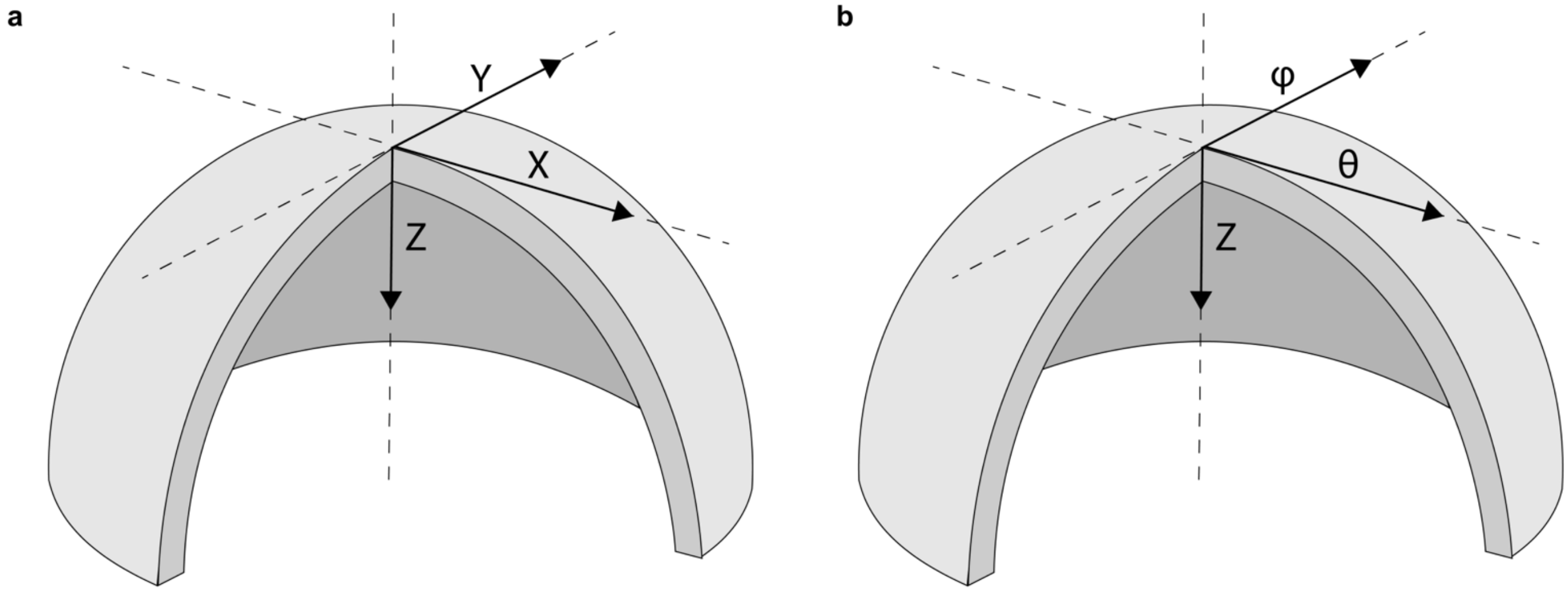

**Supplementary Fig. S2.** Corneal geometry and local coordinate axes used in analyzing deformation for (a) tensile, torsional, and wave propagation tests and (b) inflation tests.

***Hooke's law for a NITI material.*** Hooke's law relates the stress $\sigma$ and strain $\varepsilon$ in a linearly elastic solid. In its most general form, it may be written

$$\sigma_{ij} = c_{ijkl}\varepsilon_{kl}, \qquad (S2.1)$$

where summation is implied over repeated indices. In a general anisotropic solid, the fourth-rank stiffness tensor $c_{ijkl}$ contains 21 independent elastic constants. This reduces to 5 constants in a transversely isotropic solid, 3 constants in a NITI solid, and 2 constants in an isotropic solid. For convenience, Equation S2.1 is often written in Voigt notation

$$
\begin{bmatrix} \sigma_{xx} \\ \sigma_{yy} \\ \sigma_{zz} \\ \tau_{yz} \\ \tau_{xz} \\ \tau_{xy} \end{bmatrix} = \begin{bmatrix} C_{11} & C_{12} & C_{13} & C_{14} & C_{15} & C_{16} \\ * & C_{22} & C_{23} & C_{24} & C_{25} & C_{26} \\ * & * & C_{33} & C_{34} & C_{35} & C_{36} \\ * & * & * & C_{44} & C_{45} & C_{46} \\ * & * & * & * & C_{55} & C_{56} \\ * & * & * & * & * & C_{66} \end{bmatrix} \begin{bmatrix} \varepsilon_{xx} \\ \varepsilon_{yy} \\ \varepsilon_{zz} \\ \gamma_{yz} \\ \gamma_{xz} \\ \gamma_{xy} \end{bmatrix}, \qquad (S2.2)
$$

where $\tau_{ij}$ denotes shear stresses, $\gamma_{ij} = 2\varepsilon_{ij}$ denotes shear strains, and stars denote symmetric entries (i.e. $C_{21} = C_{12}$). The subscripts $x$, $y$, and $z$ refer to standard Cartesian axes, shown for the cornea in Suppl. Fig. S2a. For a NITI material, Equation S2.2 reduces to

$$
\begin{bmatrix} \sigma_{xx} \\ \sigma_{yy} \\ \sigma_{zz} \\ \tau_{yz} \\ \tau_{xz} \\ \tau_{xy} \end{bmatrix} = \begin{bmatrix} \lambda + 2\mu & \lambda & \lambda & & & \\ \lambda & \lambda + 2\mu & \lambda & & & \\ \lambda & \lambda & \lambda + 2\mu & & & \\ & & & G & & \\ & & & & G & \\ & & & & & \mu \end{bmatrix} \begin{bmatrix} \varepsilon_{xx} \\ \varepsilon_{yy} \\ \varepsilon_{zz} \\ \gamma_{yz} \\ \gamma_{xz} \\ \gamma_{xy} \end{bmatrix}. \qquad (S2.3)
$$

In the following sections, we consider the stress-strain behavior of the NITI model (Equation S2.3) under various corneal mechanical tests.

***Tensile testing of corneal strips.*** In tensile tests, rectangular strips of ex vivo cornea are subjected to uniaxial tension. The corresponding strain is measured, and the Young's modulus is quantified as $E = \sigma/\varepsilon$. In general, the orientation of the cornea strip lies in the *xy*-plane, and the direction of the applied stress aligns with the long axis of the strip. While anisotropy within the *xy*-plane has been reported by some studies, the degree of anisotropy at low IOP (low pre-stress)[6,28–31] suggest that the cornea microstructure can be approximated with the NITI model as symmetric for any direction in the *xy*-plane. Thus, it is sufficient to consider only one orientation of the strip. Consider uniaxial loading in the $x$ direction. The stress-strain relation becomes

$$
\begin{bmatrix} \sigma_{xx} \\ 0 \\ 0 \\ 0 \\ 0 \\ 0 \end{bmatrix} = \begin{bmatrix} \lambda + 2\mu & \lambda & \lambda & & & \\ \lambda & \lambda + 2\mu & \lambda & & & \\ \lambda & \lambda & \lambda + 2\mu & & & \\ & & & G & & \\ & & & & G & \\ & & & & & \mu \end{bmatrix} \begin{bmatrix} \varepsilon_{xx} \\ \varepsilon_{yy} \\ \varepsilon_{zz} \\ \gamma_{yz} \\ \gamma_{xz} \\ \gamma_{xy} \end{bmatrix}. \qquad (S2.4)
$$

It is clear from this linear system that the shear strains are zero and consequentially any equations with $G$ vanish. We are left with

$$
\begin{bmatrix} \sigma_{xx} \\ 0 \\ 0 \end{bmatrix} = \begin{bmatrix} \lambda + 2\mu & \lambda & \lambda \\ \lambda & \lambda + 2\mu & \lambda \\ \lambda & \lambda & \lambda + 2\mu \end{bmatrix} \begin{bmatrix} \varepsilon_{xx} \\ \varepsilon_{yy} \\ \varepsilon_{zz} \end{bmatrix}. \qquad (S2.5)
$$

Solving the third row for $\varepsilon_{zz}$ gives

$$
\varepsilon_{zz} = -\frac{\lambda}{\lambda + 2\mu}\left(\varepsilon_{xx} + \varepsilon_{yy}\right). \qquad (S2.6)
$$

Substituting this into the second row and solving for $\varepsilon_{yy}$ gives

$$\varepsilon_{yy} = -\frac{\lambda}{2(\lambda+\mu)}\varepsilon_{xx}. \qquad (S2.7)$$

Finally, substituting into the first row and defining the Young's modulus as $E = \sigma_{xx}/\varepsilon_{xx}$, we find

$$E = \frac{\mu(3\lambda+2\mu)}{\lambda+\mu}, \qquad (S2.8)$$

which in the incompressible limit $\lambda \to \infty$ gives $E = 3\mu$. From this, we see that a NITI material under tensile test will have an apparent Young's modulus that depends only on $\mu$ and not on $G$.

***Inflation tests of corneal trephinates.*** In corneal inflation tests, a circular region of the cornea and sclera (called a trephinate) is dissected from an ex vivo eye (see for example, Anderson et. al.[1]). It is clamped above a fluid-filled chamber and sealed around the scleral rim. The chamber is connected to a water column whose height is varied to simulate intraocular pressure (IOP). For a given pressure $p$, the test measures the rise of the corneal apex $r$. The corneal thickness $h$, radius of curvature $R$, and the contact angle at the clamped edge $\gamma$ are also estimated. Deformation of the cornea is then modeled using the theory of spherical shells.[32]

The corneal inflation test can be described by the coordinate system shown in Suppl. Fig. S2b. The deformation of an infinitesimal shell element under pressure is governed by the normal stress resultants $N_\varphi$ and $N_\theta$, bending moments $M_\varphi$ and $M_\theta$, and shear stress resultant $Q_\varphi$. Here, $\varphi$ represents the meridional coordinate (sometimes denoted by $y$ in local element coordinates) and $\theta$ denotes the azimuthal coordinate (sometimes denoted as $x$ in local element coordinates). The resultants and moments are defined:

$$N_\theta = \int_{-h/2}^{h/2} \sigma_\theta \left(1-\frac{z}{R}\right) \mathrm{d}z,$$

$$N_\varphi = \int_{-h/2}^{h/2} \sigma_\varphi \left(1-\frac{z}{R}\right) \mathrm{d}z,$$

$$M_\theta = \int_{-\frac{h}{2}}^{\frac{h}{2}} \sigma_\theta z \left(1-\frac{z}{R}\right) \mathrm{d}z, \qquad (S2.9)$$

$$M_\varphi = \int_{-h/2}^{h/2} \sigma_\varphi z \left(1-\frac{z}{R}\right) \mathrm{d}z,$$

$$Q_\varphi = \int_{-h/2}^{h/2} \tau_{\varphi z} \left(1-\frac{z}{R}\right) \mathrm{d}z.$$

Balancing the forces and moments on the shell element and considering the relationship between strain and the deformed shape leads to a system of five equations describing the rise of the corneal apex following inflation.[1]

In assigning a constitutive model to the shell, we apply Hooke's law in the local coordinate system of the shell element. As the shell is assumed to be thin relative to its radius of curvature ($h \ll R$), a plane stress approximation is also applied. This yields the following stress-strain relation for the shell element:

$$\begin{bmatrix} \sigma_\theta \\ \sigma_\varphi \\ 0 \\ 0 \\ 0 \\ \tau_{\theta\varphi} \end{bmatrix} = \begin{bmatrix} \lambda + 2\mu & \lambda & \lambda & & & \\ \lambda & \lambda + 2\mu & \lambda & & & \\ \lambda & \lambda & \lambda + 2\mu & & & \\ & & & G & & \\ & & & & G & \\ & & & & & \mu \end{bmatrix} \begin{bmatrix} \varepsilon_\theta \\ \varepsilon_\varphi \\ \varepsilon_z \\ \gamma_{\varphi z} \\ \gamma_{\theta z} \\ \gamma_{\theta\varphi} \end{bmatrix}. \tag{S2.10}$$

We immediately see, as with the tensile test, that the shear strains vanish and the solution will not depend on $G$. Proceeding as in the tensile test, we solve the third row for $\varepsilon_z$,

$$\varepsilon_z = -\frac{\lambda}{\lambda + 2\mu}(\varepsilon_\theta + \varepsilon_\varphi) \tag{S2.11}$$

and substitute into rows 1 and 2 to obtain a simplified system

$$\sigma_\theta = \frac{4\mu(\lambda + \mu)}{\lambda + 2\mu}\varepsilon_\theta + \frac{2\lambda\mu}{\lambda + 2\mu}\varepsilon_\varphi, \tag{S2.12}$$

$$\sigma_\varphi = \frac{2\lambda\mu}{\lambda + 2\mu}\varepsilon_\theta + \frac{4\mu(\lambda + \mu)}{\lambda + 2\mu}\varepsilon_\varphi. \tag{S2.13}$$

For convenience, we convert Equation S2.13 from $\lambda$-$\mu$ (Lamé parameter) form to the Young's modulus and Poisson's ratio form using the following relationships:

$$\lambda = \frac{E\nu}{(1 + \nu)(1 - 2\nu)},$$

$$\mu = \frac{E}{2(1 + \nu)}.$$

to obtain

$$\sigma_\theta = \frac{E}{1 - \nu^2}(\varepsilon_\theta + \nu\varepsilon_\varphi), \tag{S2.14}$$

$$\sigma_\varphi = \frac{E}{1 - \nu^2}(\varepsilon_\varphi + \nu\varepsilon_\theta). \tag{S2.15}$$

For an incompressible material, $\nu$ = 0.5, and the stresses depend only on the Young's modulus $E$ = 3$\mu$. As these stresses are the only terms that appear in the stress resultants and bending moments, it follows that those quantities depend only on $\mu$. The apical rise measured in an inflation test depends on these resultants and moments, and so also depends only on $\mu$. Thus, like the tensile test, an inflation test of a NITI material will measure $\mu$ and not $G$.

***Shear torsional test of corneal trephinates.*** In torsional tests of corneal trephinates, a cylindrical section of ex vivo cornea (with thickness $h$ and radius $R$) is placed between parallel platens and a rotational deformation Θ is applied to the sample. The torque $T$ at the top platen is measured, and Hooke's law for shear is used to estimate the shear modulus of the sample. The torque is given by the integral

$$T = \int_0^R \int_0^{2\pi} \tau_{\theta z} r \, \mathrm{d}\theta \mathrm{d}r \, . \tag{S2.16}$$

The transformation from Cartesian coordinates to cylindrical coordinates gives the shear stress and strain

$$\tau_{\theta z} = -\sin\theta\,\tau_{xz} + \cos\theta\,\tau_{yz}, \qquad (S2.17)$$

$$\gamma_{\theta z} = -\sin\theta\,\gamma_{xz} + \cos\theta\,\gamma_{yz}. \qquad (S2.18)$$

For the NITI material, we have

$$\tau_{xz} = G\gamma_{xz}, \qquad (S2.19)$$

$$\tau_{yz} = G\gamma_{yz}. \qquad (S2.20)$$

Substituting equations S2.19 and S2.20. into S2.17, we obtain

$$\tau_{\theta z} = G\gamma_{\theta z}. \qquad (S2.21)$$

The shear strain and stress can be approximated as

$$\gamma_{\theta z} = \frac{r\Theta}{h} \qquad (S2.22)$$

$$\tau_{\theta z} = \frac{Gr\Theta}{h} \qquad (S2.23)$$

Evaluating the torque integral and solving for $G$, we obtain

$$G = T\frac{2h}{\pi\Theta R^4}. \qquad (S2.24)$$

Clearly, the shear torsional test provides an estimate of $G$ only and is not influenced by $\mu$. Thus, modulus estimates from shear torsional tests may vary greatly from tensile and inflation measurements.

***Dynamic measurements of corneal elasticity.*** Dynamic or transient elasticity measurements use the propagation of elastic waves to estimate elastic moduli. They can provide non-destructive in vivo measurements of corneal elasticity and have the potential to be useful in clinical measurements. Because soft biological tissues are nearly incompressible, the longitudinal wave speed is large (≈1540 m/s) compared to the shear wave speed (1-10 m/s), and the shear wave speed provides a full description of tissue elasticity.

Two models of elastic wave propagation have been used to estimate corneal Young's modulus. The simplest of these treats the cornea as a semi-infinite isotropic solid.[27] This model suffers from serious inaccuracies in converting the measured group velocity to bulk shear wave speed.[33] However, we analyze the assumption here for completeness. In this case, a Rayleigh wave propagates along the air-cornea interface at constant speed proportional to the bulk shear wave speed. By measuring the group velocity of the Rayleigh wave $c_R$, one can estimate the Young's modulus as

$$E = 3\rho\left(\frac{c_R}{0.9553}\right)^2.$$

In a transversely isotropic material, the Rayleigh wave speed can be obtained numerically by employing the Stroh formalism[34–38] or evaluating the Green's function.[39] For materials with $G < \mu$, such as we expect for the cornea, the Rayleigh wave speed is primarily governed by $G$ and only slightly influenced by $\mu$ (Supplementary Note 3 provides detailed analysis of the Rayleigh wave speed in a NITI medium).

The cornea should be more accurately modeled as a bounded material. The bounded geometry gives rise to dispersive guided waves, whose frequency-wavenumber behavior must be analyzed to quantify elasticity. Partial wave analysis of the cornea as a flat isotropic plate bounded above by air and below by water leads to a secular equation that describes guided wave modes.[8,40] We detail the mode behavior (Results, Guided wave behavior in a bounded NITI layer) and the partial wave solution for a NITI material (Supplementary Note 4) elsewhere in this work.

In both our study and previous OCE studies of porcine cornea, only the $A_0$ mode is analyzed to estimate elasticity. For a bounded NITI material, the $A_0$ mode is governed primarily by $G$ and slightly influenced by $\mu$. Therefore, in both shear torsional tests and dynamic OCE tests, the observed mechanical behavior is governed primarily by $G$ rather than $\mu$ (as in tensile/inflation tests). This is consistent with shear modulus values reported by dynamic OCE studies where guided wave behavior has been taken into account through dispersion analysis of the $A_0$ mode.[8,28]

**Supplementary Note 3. Wave behavior in a bulk NITI medium**

Transversely isotropic materials support three bulk waves (quasi-longitudinal, quasi-shear, and shear) whose wave speeds depend on the propagation angle. The mechanical behavior is symmetric with respect to rotations about the *z*-axis, but varies with respect to the angle $\theta$ formed by the propagation direction and the *z*-axis (Suppl. Fig. S3.1a).

When the propagation direction lies in the *xy*-plane ($\theta$ = 90°), three pure wave modes exist (one longitudinal and two orthogonal shear waves). Their wave speeds are

$$c_L = \sqrt{C_{11}/\rho}$$

$$c_{S1} = \sqrt{C_{44}/\rho}, \text{ polarized in } z,$$

$$c_{S2} = \sqrt{C_{66}/\rho}, \text{ polarized in the } xy\text{-plane},$$

where $\rho$ is the material density.

For a general propagation angle, the polarization of propagation waves is not fully parallel or orthogonal to the propagation direction and, therefore, one quasi-longitudinal, one quasi-shear, and one shear wave should be considered. The following equations can be used to calculate their wave speeds [41]:

$$c_{qL} = \sqrt{\frac{C_{11}\sin^2\theta + C_{33}\cos^2\theta + C_{44} + \sqrt{M(\theta)}}{2\rho}}, \qquad (S2.25)$$

$$c_{qS} = \sqrt{\frac{C_{11}\sin^2\theta + C_{33}\cos^2\theta + C_{44} - \sqrt{M(\theta)}}{2\rho}},$$

$$c_S = \sqrt{\frac{C_{66}\sin^2\theta + C_{44}\cos^2\theta}{2\rho}},$$

$$M(\theta) = [(C_{11} - C_{44})\sin^2\theta + (C_{44} - C_{33})\cos^2\theta]^2 + (C_{13} + C_{44})^2\sin^2 2\theta.$$

For the NITI model, these equations simplify to

$$c_{qL} = \sqrt{\frac{\lambda + 2\mu + G + \sqrt{M(\theta)}}{2\rho}}, \qquad (S2.26)$$

$$c_{qS} = \sqrt{\frac{\lambda + 2\mu + G - \sqrt{M(\theta)}}{2\rho}},$$

$$c_S = \sqrt{\frac{\mu \sin^2\theta + G \cos^2\theta}{2\rho}},$$

$$M(\theta) = (\lambda + 2\mu - G)^2 \cos^2 2\theta + (\lambda + G)^2 \sin^2 2\theta.$$

By evaluating these equations for a range of angles using approximate mechanical properties for cornea ($\rho$ = 1000 kg/m$^3$, $\mu$ = 1 MPa, $G$ = 20 kPa, and setting $\lambda$ so that the average longitudinal wave speed is approximately 1540 m/s), we can gain some intuition of the bulk wave behavior in a NITI material. The quasi-longitudinal wave speed is nearly constant over all angles, with variations of ± 0.01%. The quasi-shear waves' speeds show a large range of variability over angle (Suppl. Fig. S3.1b).

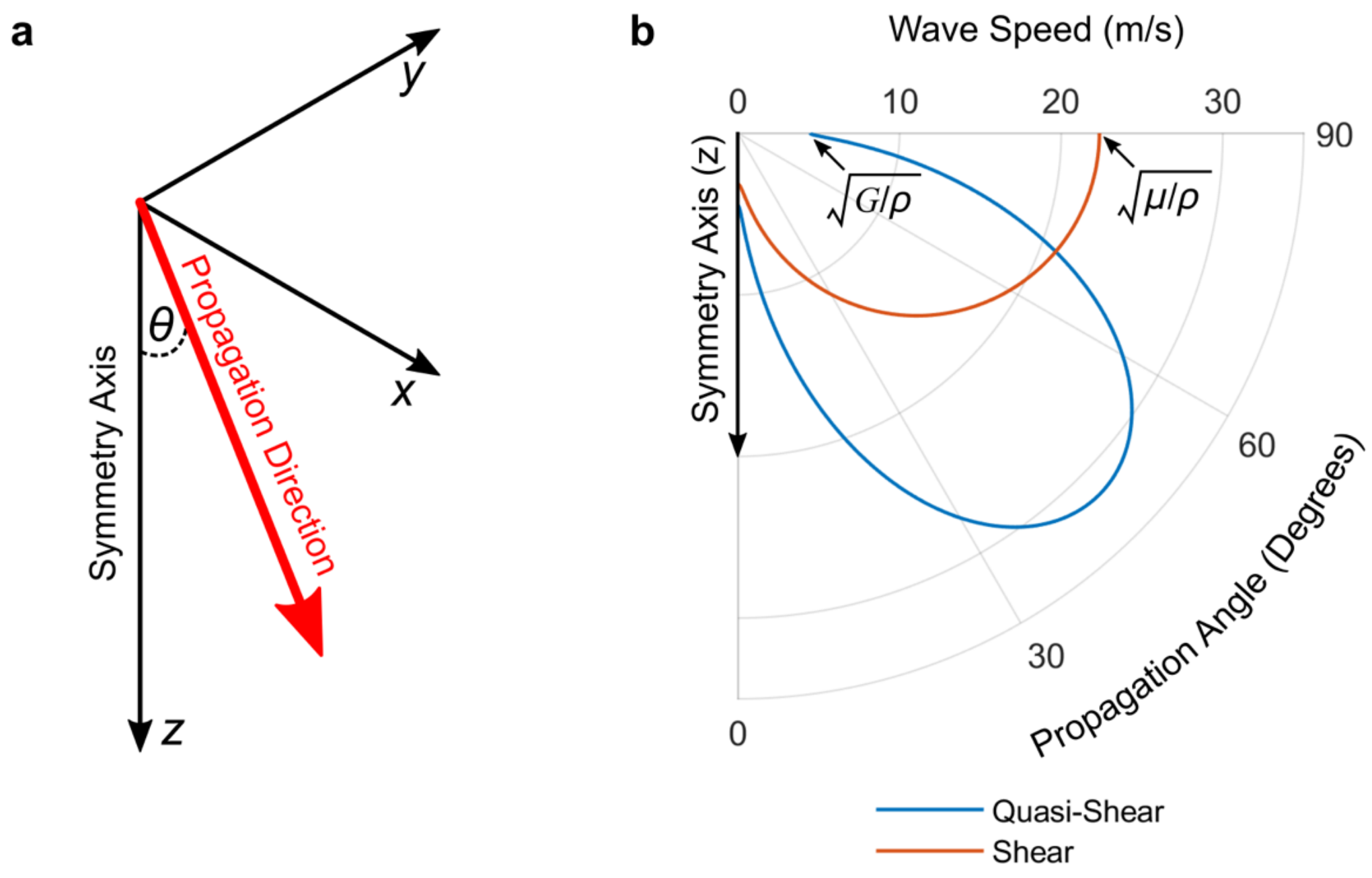

**Supplementary Fig. S3.1.** (a) Coordinate system, symmetry axis, and propagation angle for a TI material. (b) Quasi-shear and shear wave speed as a function of propagation angle for a NITI material with properties representative of cornea ($\rho$ = 1000 kg/m$^3$, $\mu$ = 1 MPa, $G$ = 20 kPa, $\lambda$ = 2.37 GPa).

When the $xy$-plane corresponds to a free surface, a Rayleigh wave propagates along the surface with speed $c_R$. The value of $c_R$ can be found using the Stroh formalism.[34–38] Here, we consider the solution of a Rayleigh wave propagating in the $\mathbf{m}$ = [1, 0, 0] direction along the free surface defined by normal vector $\mathbf{n}$ = [0, 0, 1] (because of symmetry in the $xy$-plane, the propagation direction is arbitrary). The Stroh formalism defines the 3×3 matrices

$$Q_{ik} = c_{ijkl} m_j m_l - \rho v^2 \delta_{ik} \qquad (S2.27)$$

$$R_{ik} = c_{ijkl} m_j n_l$$

$$T_{ik} = c_{ijkl} n_j n_l$$

where $c_{ijkl}$ is the fourth-order stiffness tensor, $\rho$ is the density, and $v$ is the wave speed. These matrices are combined to form the 6×6 Stroh eigenvalue problem,

$$\mathbf{N}\begin{bmatrix}\mathbf{a}\\ \mathbf{b}\end{bmatrix} = p\begin{bmatrix}\mathbf{a}\\ \mathbf{b}\end{bmatrix}, \tag{S2.28}$$

where

$$\mathbf{N} = \begin{bmatrix}-\mathbf{T}^{-1}\mathbf{R}^T & \mathbf{T}^{-1}\\ \mathbf{R}\mathbf{T}^{-1}\mathbf{R}^T - \mathbf{Q} & -\mathbf{T}^{-1}\mathbf{R}^T\end{bmatrix}.$$

The eigenvectors contain the polarization vectors $\mathbf{a}$ and traction vectors $\mathbf{b}$ of harmonic waves that satisfy the free surface boundary condition. The displacements and tractions for these solutions are

$$\mathbf{u} = \mathbf{a}e^{ik(\mathbf{m}\cdot\mathbf{x}+p\mathbf{n}\cdot\mathbf{x}-vt)}, \tag{S2.29}$$

$$\mathbf{t} = ik\mathbf{b}e^{ik(\mathbf{m}\cdot\mathbf{x}-vt)}.$$

The Rayleigh wave is formed by a linear combination of wave modes whose amplitude must decay with depth. For this reason, we only consider solutions where $p$ has a positive imaginary part. Below the limiting velocity of the material, the six eigenvalues occur in complex conjugate pairs, and so only three modes form the Rayleigh wave. Furthermore, the free surface condition implies that the linear combination of the tractions must be zero,

$$\sum_{i=1}^{3} c_i \mathbf{t}_i = \mathbf{0}. \tag{S2.30}$$

This is often rewritten in the following form:

$$ik\begin{bmatrix}| & | & |\\ \mathbf{b}_1 & \mathbf{b}_2 & \mathbf{b}_3\\ | & | & |\end{bmatrix}\begin{bmatrix}c_1\\ c_2\\ c_3\end{bmatrix}e^{ik(\mathbf{m}\cdot\mathbf{x}-vt)} = \mathbf{B}\mathbf{c} = \mathbf{0}, \tag{S2.31}$$

which has a nontrivial solution if and only if the determinant of $\mathbf{B}$ is zero.

To solve for the Rayleigh wave speed, one must employ an iterative algorithm. At each iteration, a trial wave speed $v$ is considered, and the Stroh eigenvalue problem is solved to obtain the relevant eigenvectors, as described above. The traction vectors are extracted and used to form the matrix $\mathbf{B}$ and calculate its determinant. This process is repeated until the absolute value of the determinant is minimized. The resulting wave speed corresponds to the Rayleigh wave speed $c_R$.

Following this procedure for the same representative NITI material used in the previous section ($\rho$ = 1000 kg/m$^3$, $\mu$ = 1 MPa, $G$ = 20 kPa, and setting $\lambda$ = 2.37 GPa so that the average longitudinal wave speed is approximately 1540 m/s), we found that the Rayleigh wave speed in a cornea-like NITI solid is approximately $\sqrt{G/\rho}$. The exact value varies slightly with the degree of anisotropy $G/\mu$, from approximately 0.9553$\sqrt{G/\rho}$ in the isotropic limit ($G = \mu$) to $\sqrt{G/\rho}$ in the highly anisotropic limit ($G \ll \mu$) (Fig. S3.2).

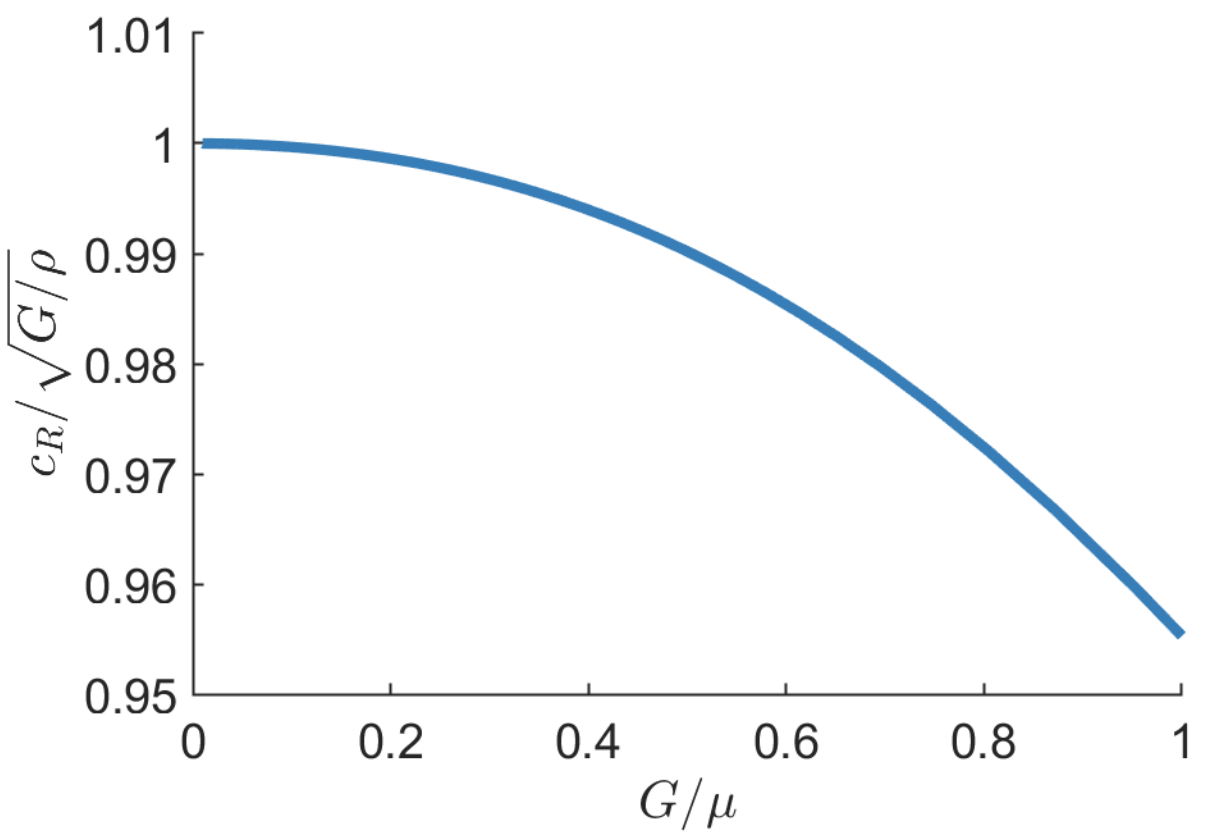

**Supplementary Fig. S3.2.** Rayleigh wave speed as a function of anisotropy. The Rayleigh wave speed is primarily governed by $G$ in a NITI solid.

**Supplementary Note 4. Analytical solution for guided wave modes in a NITI solid**

The cornea can be modeled as an infinite layer of thickness $h$ and density $\rho$ bounded above by air and below by water (Fig. S4a). Assuming a NITI material model, the stiffness tensor (Fig. S4b) contains material constants $\lambda$, $\mu$, and $G$. Because acoustic micro-tapping generates a pseudo-line source, we also assume a plane strain state.

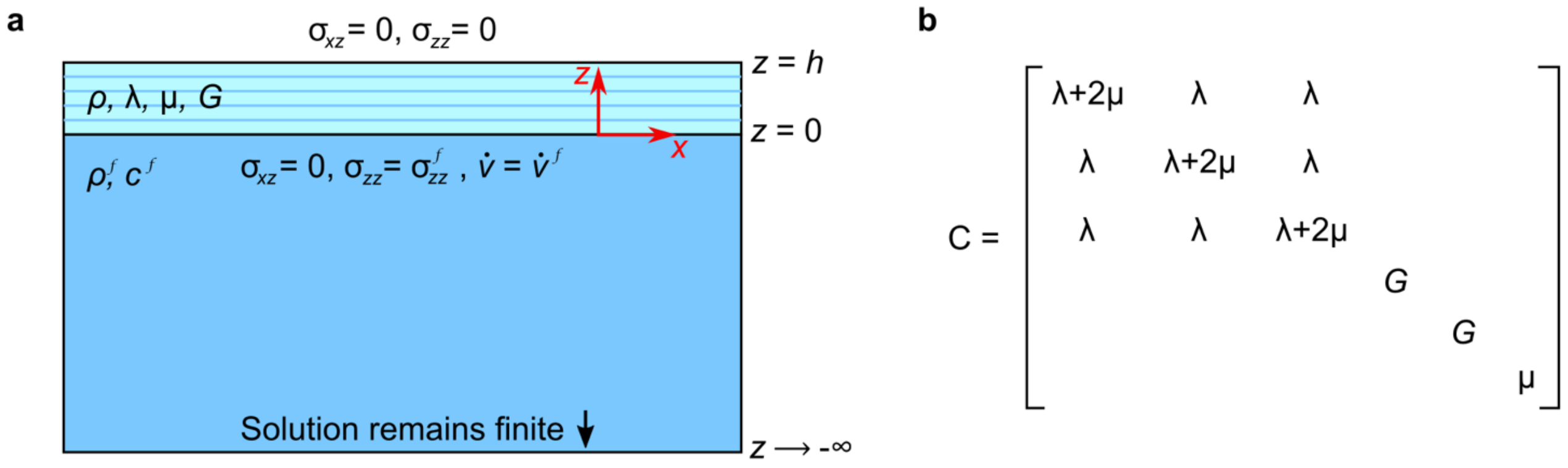

**Supplementary Fig. S4**. (a) Geometry and boundary conditions and (b) stiffness tensor for a NITI plate bounded above by air and below by water.

Introducing scales

Position: $\boldsymbol{x} \sim h$

Displacement: $\boldsymbol{u} \sim h$

Time: $t \sim h/\sqrt{\mu/\rho}$

Frequency: $f \sim \sqrt{\mu/\rho}/h$

Wavenumber: $k \sim h$

the dimensionless equations of motion yield the elastodynamic equations for a NITI material

$$u_{tt} = \beta^2 u_{xx} + \alpha^2 u_{zz} + \gamma^2 v_{xz} \tag{S4.1}$$

$$v_{tt} = \alpha^2 v_{xx} + \beta^2 v_{zz} + \gamma^2 u_{xz} \tag{S4.2}$$

$$\alpha^2 = \frac{G}{\mu}$$

$$\beta^2 = \frac{\lambda + 2\mu}{\mu}$$

$$\gamma^2 = \frac{\lambda + G}{\mu}$$

Here, $\boldsymbol{u} = (u, v)$ is the dimensionless displacement field with $x$ and $z$ components, respectively; $\alpha$, $\beta$, and $\gamma$ are dimensionless parameters, and subscripts denote partial differentiation. We assume harmonic solutions for the displacements of the form

$$u(x, z, t) = Ae^{i(kx+\ell z-\omega t)} \tag{S4.3}$$

$$v(x, z, t) = Be^{i(kx+\ell z-\omega t)} \tag{S4.4}$$

where $k$ and $\ell$ are dimensionless angular wavenumbers, $\omega$ is the dimensionless angular frequency, and $A$ and $B$ are arbitrary constants.

Substituting these solutions into the governing equations yields the homogeneous linear system

$$\begin{bmatrix} q_\beta^2 + \frac{\alpha^2}{\beta^2}\ell^2 & k\ell\frac{\gamma^2}{\beta^2} \\ k\ell\frac{\gamma^2}{\alpha^2} & q_\alpha^2 + \frac{\beta^2}{\alpha^2}\ell^2 \end{bmatrix} \begin{bmatrix} A \\ B \end{bmatrix} = \begin{bmatrix} 0 \\ 0 \end{bmatrix} \tag{S4.5}$$

where

$$q_\alpha^2 = k^2 - \frac{\omega^2}{\alpha^2}$$

$$q_\beta^2 = k^2 - \frac{\omega^2}{\beta^2}.$$

This has non-trivial solutions if and only if the determinant of the matrix is zero, resulting in a biquadratic equation for $\ell$ with four solutions

$$\ell^4 + \left[\frac{\alpha^2}{\beta^2}q_\alpha^2 + \frac{\beta^2}{\alpha^2}q_\beta^2 - \frac{\gamma^4 k^2}{\alpha^2\beta^2}\right]\ell^2 + q_\alpha^2 q_\beta^2 = 0 \tag{S4.6}$$

$$\ell = \pm\sqrt{\frac{1}{2}\left[\varphi \pm \sqrt{\varphi^2 - 4q_\alpha^2 q_\beta^2}\right]} \tag{S4.7}$$

$$\varphi = \frac{\gamma^4 k^2}{\alpha^2\beta^2} - \frac{\alpha^2}{\beta^2}q_\alpha^2 - \frac{\beta^2}{\alpha^2}q_\beta^2$$

Without loss of generality, we assume $B$ = 1 and solve the corresponding coefficient $A$ for each solution $\ell$. This yields the following solutions, where the outer and inner ± signs match for each $A$ and $\ell$,

$$A = \pm\left[-\frac{\sqrt{2}\frac{\gamma^2 k}{\alpha^2}\sqrt{\varphi \pm \sqrt{\varphi^2 - 4q_\alpha^2 q_\beta^2}}}{\varphi + 2\frac{\beta^2}{\alpha^2}q_\beta^2 \pm \sqrt{\varphi^2 - 4q_\alpha^2 q_\beta^2}}\right] \qquad (S4.8)$$

The full solutions are therefore linear combinations of four partial waves,

$$u(x,z,t) = \sum_{j=1}^{4} C_j A_j e^{i\ell_j z} e^{i(kx-\omega t)} \qquad (S4.9)$$

$$v(x,z,t) = \sum_{j=1}^{4} C_j e^{i\ell_j z} e^{i(kx-\omega t)} \qquad (S4.10)$$

In the fluid domain, we assume an acoustic material. The dimensionless acoustic wave equation in velocity potential form is

$$\dot{\boldsymbol{u}}^f = \nabla\Phi \qquad (S4.11)$$

$$p^f = -\frac{\rho^f}{\rho}\Phi_t \qquad (S4.12)$$

$$\Delta\Phi - \frac{1}{\delta^2}\Phi_{tt} = 0 \qquad (S4.13)$$

$$\delta^2 = \frac{\rho\left(c^f\right)^2}{\mu}$$

The general solution for the fluid domain takes the form

$$\Phi = C_5 e^{\xi z} e^{i(kx-\omega t)} \qquad (S4.14)$$

$$\xi = \sqrt{k^2 - \frac{\omega^2}{\delta^2}}$$

$$\mathrm{Re}(\xi) > 0$$

The constants $C_j$ are chosen so that the solutions satisfy the non-dimensionalized boundary conditions:

$$\begin{aligned} \sigma_{xz} &= 0 \quad \text{at } z = 1 \\ \sigma_{zz} &= 0 \quad \text{at } z = 1 \\ \sigma_{xz} &= 0 \quad \text{at } z = 0 \\ \sigma_{zz} &= \sigma_{zz}^f \quad \text{at } z = 0 \\ \dot{v} &= \dot{v}^f \quad \text{at } z = 0 \end{aligned} \qquad (S4.15)$$

where the superscripts $f$ denote quantities in the fluid domain $z < 0$. Substituting the general solutions into the boundary conditions yields a 5×5 homogeneous system for the coefficients, $\mathbf{Mc} = \mathbf{0}$. This system has nontrivial solutions if and only if the determinant of the matrix $\mathbf{M}$ (equation S4.16) is zero. For a given frequency $\omega$, wavenumbers $k$ can be found that satisfy the dispersion relation using numerical root-finding methods or by minimizing the absolute value of the determinant

(S4.16)

$$\mathbf{M} = \begin{bmatrix} (\ell_1 A_1 + k)e^{i\ell_1} & (\ell_2 A_2 + k)e^{i\ell_2} & (\ell_3 A_3 + k)e^{i\ell_3} & (\ell_4 A_4 + k)e^{i\ell_4} & 0 \\ [k(\gamma^2 - \alpha^2)A_1 + \beta^2 \ell_1]e^{i\ell_1} & [k(\gamma^2 - \alpha^2)A_2 + \beta^2 \ell_2]e^{i\ell_2} & [k(\gamma^2 - \alpha^2)A_3 + \beta^2 \ell_3]e^{i\ell_3} & [k(\gamma^2 - \alpha^2)A_4 + \beta^2 \ell_4]e^{i\ell_4} & 0 \\ \ell_1 A_1 + k & \ell_2 A_2 + k & \ell_3 A_3 + k & \ell_4 A_4 + k & 0 \\ k(\gamma^2 - \alpha^2)A_1 + \beta^2 \ell_1 & k(\gamma^2 - \alpha^2)A_2 + \beta^2 \ell_2 & k(\gamma^2 - \alpha^2)A_3 + \beta^2 \ell_3 & k(\gamma^2 - \alpha^2)A_4 + \beta^2 \ell_4 & \omega \rho^f / \rho \\ \omega & \omega & \omega & \omega & -i\xi \end{bmatrix}$$

**Supplementary Note 5. Finite element model of a NITI cornea**

To provide an ideal, noise-free comparison to our experimental measurements, we developed a two-dimensional (plane strain) finite element model of guided wave propagation in a NITI material using OnScale (OnScale, Redwood City, CA). The model mirrors the configuration of our OCE experiments in porcine cornea. A thin elastic layer of thickness $h$ = 0.55 mm and density $\rho$ = 1000 kg/m$^3$ is bounded above by air (free surface condition) and below by a layer of water (Fig. S5a). The elastic layer is modeled as a NITI material with parameters $\lambda$, $\mu$, and $G$, while the water layer is modeled as an isotropic solid with a density $\rho^f$ = 1000 kg/m$^3$, shear wave speed $c_s = 0$, and longitudinal wave speed $c_L = \sqrt{(\lambda + 2\mu)/\rho}$.

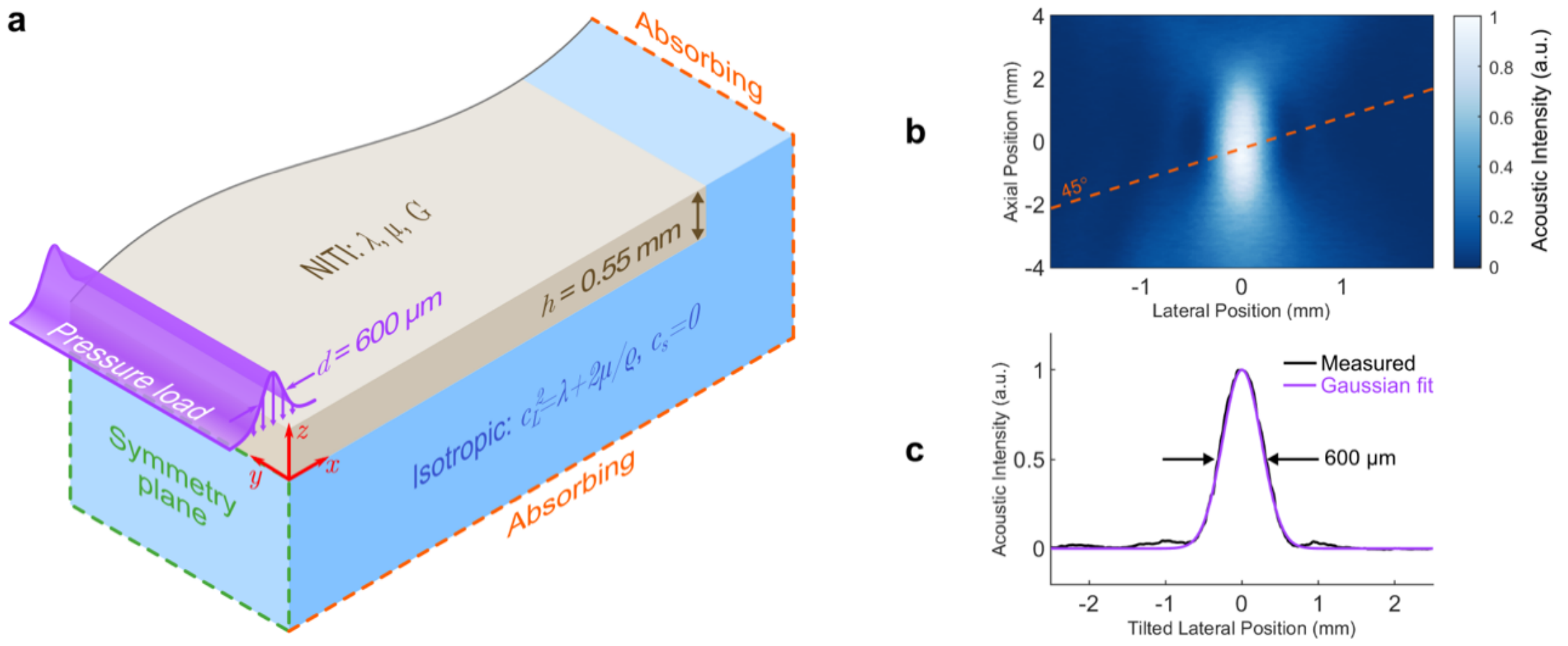

**Supplementary Fig. S5**. (a) Finite element model geometry used to simulate guided wave propagation in a NITI model of the cornea. (b) Acoustic intensity field measured from the air-coupled AμT transducer shows a nearly Gaussian profile (c) when sampled along a 45° line through the transducer focus.

The bottom and right boundaries of the domain function as absorbing conditions to minimize reflections as elastic waves leave the computational domain. To ensure stability at the boundary, the water layer extends along the right edge of the tissue domain and maintains the boundary condition. Symmetry is assumed with respect to the $yz$-plane. A temporally- and spatially-varying pressure load $P(x,t)$ is applied to the top boundary to mimic the AμT push of OCE experiments. The spatial profile is modeled as a Gaussian with a full-width-at-half-max (FWHM) $d$ = 600 μm. This value was obtained by measuring the ultrasonic field of the AμT transducer with a needle hydrophone (HNC-1000, Onda, Sunnyvale, CA, USA) in air, sampling along a 45° line passing through the transducer focus (Fig. S5b and S5c). The temporal push profile is modeled by a super-Gaussian function with FWHM $T$ = 100 μs. A small offset $t_0$ is included to avoid impulsive loading at $t$ = 0. The full form of the pressure load is given by

$$P(x,t) = P_0 \exp\left[-4(\ln 2)\left(\frac{x}{d}\right)^2\right] \exp\left[-16(\ln 2)\left(\frac{t-t_0}{T}\right)^4\right],$$

where the pressure amplitude $P_0$ is an arbitrary constant (5 kPa in this study).

Biological tissue is nearly incompressible, exhibiting longitudinal wave speeds on the order of 1540 m/s and shear wave speeds in the range of 1-10 m/s. This corresponds to a Poisson's ratio of $\nu \approx 0.4999995$. In practice, it is not necessary to directly model this exact value of Poisson's ratio, and solutions for $\nu >$ 0.4995 converge to the incompressible solution.[33] In this study, we enforce this constraint for all models by requiring that the ratio $\sqrt{(\lambda + 2\mu)/\mu}$ = 35. In the isotropic limit $G = \mu$, this gives a Poisson's ratio of $\nu$ = 0.4996.

The computational domain was discretized using linear finite elements on a regular rectangular grid with a minimum of 40 elements per elastic wavelength, as approximated by the push width. Belytchko-Bindeman hourglass suppression was applied to prevent spurious modes from corrupting the solution. All simulations used an explicit time-stepping method to generate the full displacement and velocity fields at each space and time point. However, only the vertical velocity component was used in our analysis. This is analogous to OCE experiments where only this component is available.

**Supplemental Note 6. Finite element model to investigate the effect of corneal curvature**

In the analytical solution described in Supplementary Note 4 and the finite element model described in Supplementary Note 5, we model the cornea as a flat NITI layer bounded above by air and below by water. In reality, the cornea is not flat, but rather curved, and this curvature could affect guided wave behavior. We developed a finite element model in OnScale to estimate the effect of curvature on the dispersion behavior. This model was similar to that described in Supplementary Note 5, but featured a few distinct differences in both geometry and post-processing.

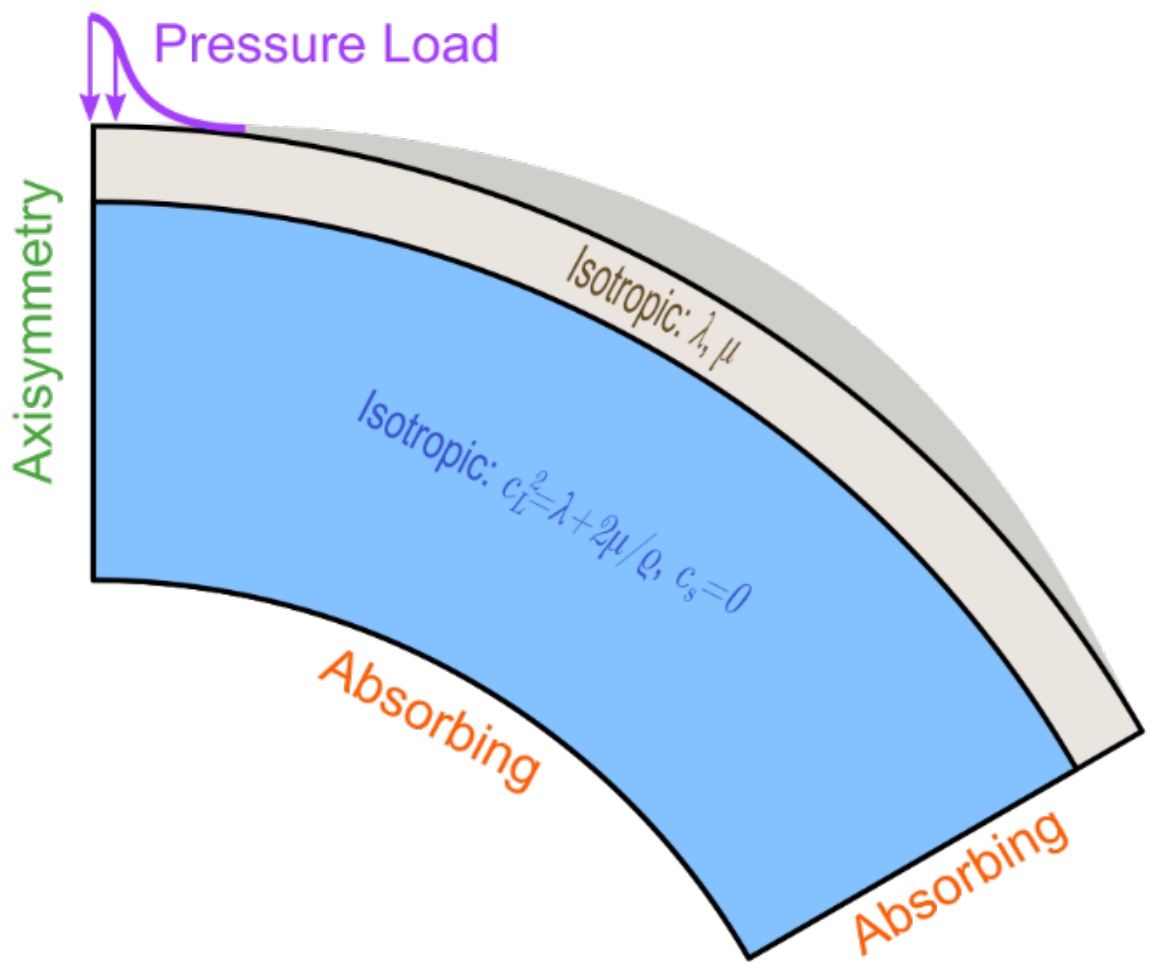

**Supplementary Fig. S6.1.** Curved geometry used to study guided waves in a spherically curved layer. The full model geometry can be considered a solid of revolution obtained by revolving the domain shown about the axisymmetric boundary.

The finite element model consisted of a rectangular domain of thickness $h$ = 0.55 mm bent to an outer radius of curvature of 6.5 mm (Supplementary Fig. S6.1). Here we assume an axisymmetric rather than a plane strain state. The tissue layer was assumed to be isotropic with density $\rho$ = 1000 kg/m$^3$, shear wave speed $c_S$ = 4 m/s, and longitudinal wave speed $c_L$ = 35$c_S$. The tissue layer sat atop a simulated water layer, modeled as an elastic solid with shear wave speed $c_S$ = 0 m/s and a density and longitudinal wave speed matched to the tissue layer. Outer edges of the domain were set as absorbing boundaries. The spatio-temporal push profile was defined as in Supplemental Note 5, with a Gaussian profile in space and a super-Gaussian profile in time. The domain was discretized with linear finite elements on a regular conformal grid with at least 40 elements per elastic wavelength.

Solving this model produces a pseudo-point source solution, a numerical approximation to the Green's function for this problem. The wave field solution for a line source, such as the one used in our AμT experiments, can be obtained by convolving the Green's function with a source distribution. We numerically evaluated this convolution integral along a linear path on the spherical surface using trapezoid integration. This yielded the wave field along the entire spherical surface. Supplemental Video 1 shows the results. In general, the wave field is more complex compared to the flat plate plane strain approximation. However, extracting the wave field from the mid-plane normal to the line source (as would be measured using OCE), we find that the flat plate and spherical models produce very similar wave fields (Supplementary Fig. S6.2, a-b). Analyzing the 2D Fourier spectrum of the spherical model along this mid-plane, we find that the isotropic dispersion relation for a flat plate[40] provides a close fit to the numerical solution (Supplementary Fig. S6.2, c). This is expected, as the radius of curvature of the cornea is large relative to its thickness.[42] Thus, it appears reasonable to neglect corneal curvature in analyzing guided wave behavior in the cornea. While these computations were performed using an isotropic model, we expect a NITI model to produce similar results.

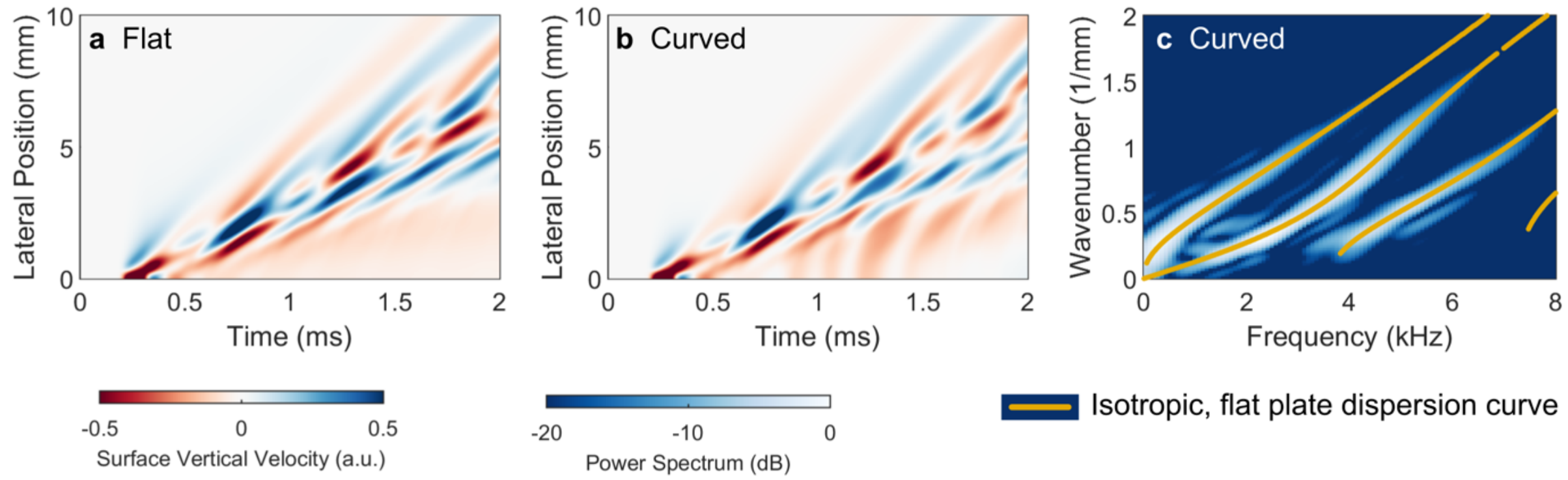

**Supplementary Fig. S6.2.** Surface vertical velocity wave fields for a simulated (a) flat plate and (b) spherically curved layer are very similar. (c) The 2D Fourier spectrum for the spherically curved layer closely matches the theoretical dispersion curves obtained for a flat plate (yellow markers).